\documentclass[a4paper,11pt]{article}
\pdfoutput=1 % if your are submitting a pdflatex (i.e. if you have
\usepackage{jheppub} % for details on the use of the package, please
\usepackage{verbatim}

\usepackage{graphicx}	% Include figure files
\usepackage{dcolumn}	% Align table columns on decimal point
\usepackage{bm}			% bold math
\usepackage{hyperref}
\usepackage{amsmath}
\usepackage{verbatim}
\usepackage{slashed}
\usepackage{hyperref}
\usepackage{mathtools}
\usepackage{tabularx}

\title{\boldmath Explaining the $R_{K^{(*)}}$ anomalies in a Fundamental Composite Higgs Model with Gauged $U(1)_{SM_3-HB}$}

\author{Yi Chung}

\affiliation{Center for Quantum Mathematics and Physics (QMAP), Department of Physics, \\University of California,  Davis, CA 95616, U.S.A.}

\emailAdd{yichung@ucdavis.edu}

\abstract{A new heavy $Z'$ vector boson provides a possible explanation for the neutral current B anomalies. Various $U(1)'$ gauge groups have been proposed and studied. In this paper, we explore a new type of $U(1)'$ gauge symmetry inspired by fundamental composite Higgs models with hyperfermions. It is also the first attempt to connect such a $Z'$ boson with a solution to the hierarchy problem. The $U(1)'$ symmetry is identified as the quantum number of the difference between the Standard Model fermion number and hyperbaryon number, written as $SM-HB$. This type of $U(1)'$ gauge symmetry is naturally broken in some fundamental composite Higgs models, which leads to a TeV-scale $Z'$ boson. We present a concrete example based on the minimal fundamental composite Higgs model with $SU(4)/Sp(4)$ coset, where the $Z'$ boson is the only state below the compositeness scale beside the composite Higgs. We also show that if the $U(1)'$ symmetry is third-generation-philic, written as $SM_3-HB$, the corresponding $Z'$ boson can explain the neutral current B anomalies. The model introduces the composite Higgs and the $Z'$ boson from the same symmetry breaking scale, so Higgs physics and flavor physics are now connected. After considering all the experimental constraints and theoretical preferences, we found that there is still a natural parameter space for $SU(4)/Sp(4)$ fundamental composite Higgs model with $N_{HC}=2$, which can be probed in the near future.}

\begin{document}
\maketitle
\flushbottom

%\newpage
\section{Introduction}

%~\\- Standard Model ~\\

The Standard Model (SM) of particle physics successfully describes all known elementary particles and their interactions. With the discovery of light Higgs bosons in 2012 \cite{Chatrchyan:2012xdj, Aad:2012tfa}, the last missing piece of the SM seemed to be filled. However, the SM does not address the UV-sensitive nature of scalar bosons. The naturalness principle requires some new physics not too far from the electroweak scale. Many models aiming to solve the hierarchy problem have predicted new particles with various properties and phenomenology around the TeV scale. But yet, none of them have been observed, which makes the hierarchy problem more challenging and profound.

%~\\ - B Anomalies ~\\

Although the direct searches by ATLAS and CMS have not provided any evidence of new particles, a set of interesting discrepancies emerged from the precise measurements of B-meson semileptonic decays by LHCb. For example, the measurement of lepton flavor universality (LFU) ratio $R(K)\equiv BR(B\to K\mu^+\mu^-)/BR(B\to Ke^+e^-)$ shows growing hints of beyond the Standard Model (BSM) physics \cite{LHCb:2014vgu, LHCb:2019hip, LHCb:2021trn}. Other related measurements include $R(K^*)$ \cite{LHCb:2017avl} and angular observables of the related decay \cite{LHCb:2013ghj, LHCb:2015svh, LHCb:2020lmf}. Each anomaly is not statistically significant enough to reach the discovery level, but the combined analysis shows a consistent deviation from the SM prediction \cite{Altmannshofer:2021qrr, Cornella:2021sby, Geng:2021nhg, Alok:2019ufo, Alguero:2021anc, Carvunis:2021jga, Hurth:2021nsi}. They are known as neutral current B anomalies (NCBAs). The global fit results point to the additional contribution to the operators
\begin{equation}
\Delta\mathcal{L}_{\text{eff}}=\frac{4G_F}{\sqrt{2}}V_{tb}V_{ts}^*\frac{e^2}{16\pi^2}
~C_{9(10)}(\bar{s}\gamma_\mu P_Lb)(\bar{\mu}\gamma^\mu(\gamma^5)\mu)
\quad\text{with}\quad |C_{9(10)}|\sim 1~,
\end{equation}
which implies a generic scale of new physics $\Lambda_{\text{NP}}\sim 36$ TeV. However, if the couplings receive CKM-like suppression, the scale can be brought down to a few TeV, which is where we now expect a solution to the hierarchy problem! The main goal of this paper is to find the most economical model to realize this idea.

Among all kinds of possible explanations for the NCBAs, the two tree-level mediators, Leptoquarks and $Z'$ bosons, are the most popular choices. Leptoquarks, however, are not the common ingredient in the solutions to the hierarchy problem. Some efforts have been made to connect leptoquarks to the electroweak scale \cite{Gripaios:2014tna, Barbieri:2016las, Blanke:2018sro, Marzocca:2018wcf, Fuentes-Martin:2020bnh}, but they all require some unusual extension to connect quarks and leptons. On the other hand, $Z'$ bosons seem to be a more natural choice. A $Z'$ boson is a massive gauge boson of some broken $U(1)'$ gauge symmetry. The $U(1)'$ symmetry can be a remnant of some non-abelian gauge groups, which is common in the extension of the SM. The real question is how to break this symmetry around the TeV scale. From another point of view, it requires a new Goldstone boson, which becomes the longitudinal degree of freedom of the $Z'$. Based on this requirement, the best candidate turns out to be a Composite Higgs Model (CHM), where additional Goldstone bosons are usually introduced.

%~\\ - Composite Higgs Model ~\\

In a CHM, the Higgs doublet is the pseudo-Nambu-Goldstone boson (pNGB) of a spontaneously broken global symmetry of the underlying strong dynamics~\cite{Kaplan:1983fs, Kaplan:1983sm}. Through the analogy to the chiral symmetry breaking in quantum chromodynamics (QCD), which naturally introduces light scalar fields, i.e., pions, we can construct models with light Higgs bosons in a similar way. In a CHM, an approximate global symmetry $G$ is spontaneously broken by some strong dynamics down to a subgroup $H$ at a symmetry breaking scale $f$. The heavy resonances of the strong dynamics are expected to be around the compositeness scale $\sim 4\pi f$ generically. The pNGBs of the symmetry breaking, on the other hand, can naturally be light with masses below $f$ as they are protected by shift symmetry.

Among all types of CHMs with various cosets, the CHMs with fundamental gauge dynamics featuring only fermionic matter fields are of interest in many studies \cite{Barnard:2013zea, Ferretti:2013kya, Cacciapaglia:2014uja, Cacciapaglia:2020kgq}, which is known as the fundamental composite Higgs model (FCHM). In this type of CHMs, hyperfermions $\psi$ are introduced as the representation of hypercolor (HC) group $G_{HC}$. Once the HC group becomes strongly coupled, hyperfermions form a condensate, which breaks the global symmetry $G$ down to the subgroup $H$. In FCHMs, there are always more than four pNGBs required by the SM scalar sector.\footnote{The opposite is Minimal Composite Higgs Models~\cite{Agashe:2004rs} with the symmetry breaking $SO(5) \to SO(4)$, which contain exactly four pNGBs. However, this symmetry breaking pattern can not be realized with fundamental fermionic matter theory naturally.} These states used to be the shortcomings of the models because they are expected to be light and within the searches of LHC, which are strongly constrained. However, if the corresponding symmetries are gauged, they can provide what we need to explain the NCBAs. Especially, the minimal FCHM, which is based on the $SU(4)/Sp(4)$ coset \cite{Katz:2005au, Gripaios:2009pe, Galloway:2010bp}, contains only five pNGBs. The four of them formes the SM Higgs doublet, and the fifth one, as a SM singlet, can give us a TeV-scale $Z'$ boson if the corresponding $U(1)'$ symmetry is local. With only one additional pNGB, the $SU(4)/Sp(4)$ FCHM could be the most economical solution to the NCBAs and the hierarchy problem.

%~\\ - $U(1)'$ symmetry ~\\

The immediate question to ask is: What is the corresponding $U(1)'$ gauge symmetry? It turns out that it is the baryon number of hyperfermions, i.e., hyperbayron number (HB). To make the $U(1)'$ symmetry free from the quantum anomaly, there should be additional fermions charged under it. The SM fermions can take this responsibility! The resulting symmetry becomes $U(1)_{SM-HB}$, where $SM$ stands for the SM fermion number. It is like a hyperversion of anomaly-free $U(1)_{B-L}$ symmetry. This anomaly-free $U(1)'$ symmetry opens a new category of $U(1)'$ symmetry which can provide a novel $Z'$ explanation to the NCBAs. In this paper, we study the $U(1)_{SM_3-HB}$ symmetry where only the third generation SM fermions are charged. We will show the corresponding $Z'$ boson can provide a good explanation to the NCBAs. A detailed study of indirect and direct experimental constraints will be presented in this study.

%~\\ - Model description ~\\

\begin{figure}[t]
\centering
\includegraphics[width=0.9\linewidth]{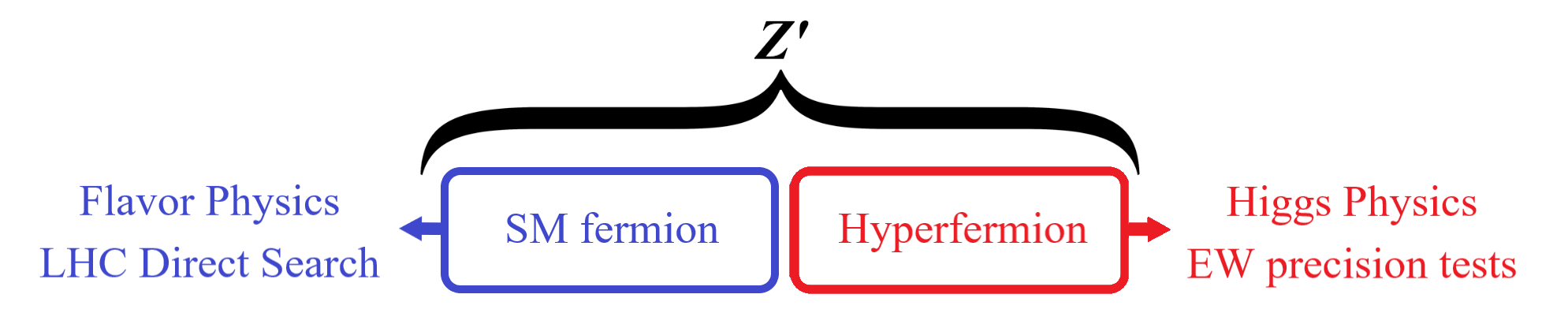}
\caption{The $Z'$ boson couples to both the SM sector and the hypercolor sector, which results in a rich phenomenology. On the left-hand side, its couplings with the SM fermions will lead to deviations in flavor physics and signatures in collider searches. On the right-hand side, the $Z'$ boson gets its mass from the same scale as composite Higgs, so the scale is connected with the Higgs and electroweak measurements.}
\label{Zboson}
\end{figure}

Combining all the ingredients, we end up with a $SU(4)/Sp(4)$ FCHM with gauged $U(1)_{SM_3-HB}$ symmetry, which we claim is the most economical model to solve all the problems. That also means the $Z'$ boson appears in all the SM phenomenology, which makes it a very predictive model (see figure~\ref{Zboson}). The scale of the $Z'$ boson comes from the hyperfermion condensate, which is the same as the composite Higgs. Therefore, we can get the constraint on the $Z'$ scale from the Higgs and electroweak measurements. The theoretical issue in CHMs will also play a role in determining the favored parameter space. The couplings of $Z'$ are restricted to the third generation SM fermions. However, it is the third generation in the flavor basis, which is not necessary to be aligned with the third generation in the mass basis. Therefore, additional contributions on flavor changing neutral currents (FCNCs) are expected. On the one hand, it can naturally explain the observed NCBAs. On the other hand, it is strongly constrained by other FCNC measurements. Besides the flavor physics part, coupling with the SM fermions also allows us to search it in the LHC, which gives us a direct bound on its mass $M_{Z'}$. In the end of this paper, we will combine all the experimental constraints and theoretical preferences.

%~\\ - Organization of the paper ~\\

This paper is organized as follows. In section~\ref{sec:U(1)'}, we introduce the $U(1)'$ symmetry which naturally exists in FCHMs. Its properties are discussed in detail. A concrete example based on $SU(4)/Sp(4)$ FCHM with its fermionic UV completion is also introduced. The $U(1)_{SM_3-HB}$ gauge symmetry, which is the difference between the third generation SM fermion number and the hyperbaryon number, was chosen to study in this paper. Next, in section~\ref{sec:Model}, we calculate the properties of $SU(4)/Sp(4)$ FCHM through the nonlinear sigma model. The gauge sector, including the SM gauge group with an additional $U(1)'$, and the Yukawa sector are presented. The loop-induced Higgs potential and the fine-tuning are also discussed. The analyses from Higgs and electroweak precision tests are shown at the end of this section. Next, we turn to the $Z'$ phenomenology in section~\ref{sec:Z'}. Start with the specified couplings to the SM fermions. We then discuss the flavor physics in section~\ref{sec:Flavor}, including the B anomalies and other experimental constraints. Direct searches are discussed in section~\ref{sec:Collider}, which play an important role in constraining a TeV-scale $Z'$ boson. The ultimate analysis combining the requirement all the way from Higgs physics to flavor physics is shown in section~\ref{sec:Combine}. Section~\ref{sec:Conclusion} contains our discussions and conclusions. In Appendix \ref{sec:Origin}, we describe the possible UV origins of the $U(1)'$ gauge symmetry.

%\newpage

\section{The $U(1)'$ gauge symmetry in fundamental composite Higgs models}\label{sec:U(1)'}

Various $U(1)'$ gauge symmetry have been proposed to explain the NCBAs \cite{Altmannshofer:2014cfa, Altmannshofer:2015mqa, Altmannshofer:2019xda, Crivellin:2015mga, Crivellin:2016ejn, Alonso:2017bff, Alonso:2017uky, Bonilla:2017lsq, Allanach:2020kss, Allanach:2018lvl, Allanach:2019iiy, Gauld:2013qba, Buras:2013dea, Buras:2013qja, King:2017anf, King:2018fcg, AristizabalSierra:2015vqb, Celis:2015ara, Falkowski:2015zwa, Chiang:2016qov, Boucenna:2016wpr, Boucenna:2016qad, Bhatia:2017tgo, Ko:2017lzd, Tang:2017gkz, Fuyuto:2017sys, Bian:2017xzg, Duan:2018akc, Calibbi:2019lvs}. Some of the models are more ambitious than others, trying to address problems such as the flavor puzzle. However, none of them tries to connect it with the hierarchy problem. This study, following the previous one \cite{Chung:2021ekz}, is the first attempt of this type.\footnote{In this work, more details of CHMs, including a general hypercolor group $SP(N_{HC})$ and the Higgs potential, are studied and the constraints on the models from Higgs and electroweak sector are discussed. We also analyze the parameter space in the small angle region where the phenomenology is very different from the previous study.} The most important idea is bringing up a new kind of $U(1)'$ symmetry inspired by FCHMs. In this section, we will explain what this $U(1)'$ symmetry is, why it is naturally broken, and how it could be gauged without quantum anomaly.

\subsection{Broken $U(1)_{HB}$ symmetry in FCHMs}

In fundamental composite Higgs models, additional hyperfermions $\psi$ are added to generate composite Higgs as their bound states. Hyperfermions are representations of hypercolor group, which becomes strongly coupled around the TeV scale. The resulting hyperfermion condensate breaks the global symmetry of hyperfermions and introduces the composite Higgs doublet as the pNGBs of the broken global symmetry. The SM electroweak symmetry is broken subsequently after the Higgs gets a VEV, which is called vacuum misalignment. In other words, hyperfermions can form two types of vacuums, electroweak conserving vacuum $\Sigma_{EW}$ and electroweak breaking vacuum $\Sigma_\slashed{EW}$. The true vacuum is the linear combination of the two vacua but is dominated by the electroweak conserving one.

These vacua can also be understood through the mass term they generate. Assume the hyperfermions have a similar chiral structure as the SM fermions.\footnote{The similar chiral structure means that the left-handed hyperfermions are the representations of $SU(2)_L\,(SU(2)_W)$ and the right-handed hyperfermions are the representations of $SU(2)_R\,(U(1)_Y)$. It is common in FCHMs based on $SU(N)/SO(N)$ and $SU(N)/Sp(N)$. However, it might not be the case for $SU(N)\times SU(N)/SU(N)$ FCHMs.} The electroweak breaking vacuum gives a Dirac mass term because it is the same as the chiral symmetry breaking vacuum. The mass term by the electroweak conserving vacuum is similar to the Majonara mass term. Similar to the Majorana mass term of the right-handed neutrino, which breaks the lepton number conservation, the electroweak conserving vacuum also breaks a quantum number in FCHMs. It was called hyperbaryon number $HB$, which is defined as
\begin{equation}
Q_{HB}\equiv\frac{1}{N_{HC}}\left(n_\psi-n_{\bar{\psi}}\right),
\end{equation}
where $N_{HC}$ is the number of hypercolor, $n_\psi$ is the number of hyperfermions, and $n_{\bar{\psi}}$ is the number of anti-hyperfermions.

A concrete example is based on the minimal fundamental composite Higgs model with the global symmetry breaking $SU(4)\to Sp(4)$ \cite{Cacciapaglia:2014uja, Cacciapaglia:2020kgq}. The fermionic UV completion of a $SU(4)/Sp(4)$ FCHM only requires 4 Weyl fermions in the fundamental representation of the $Sp(N_{HC})$ hypercolor group, where $N_{HC}$ is even. The 4 Weyl hyperfermions transform under $G_{SM}=SU(3)_C\times SU(2)_W\times U(1)_Y$ as
\begin{align}
(U_L,D_L)=(1,2)_0~,\quad U_R=(1,1)_{1/2}~,\quad D_R=(1,1)_{-{1/2}}~,
\end{align}
which is similar to the SM fermions. After rewriting the two right-handed hyperfermions as $\tilde{U}_L=-i\sigma_2C\bar{U}^T_R$ and $\tilde{D}_L=-i\sigma_2C\bar{D}^T_R$, the 4 Weyl fermions are now according to the same representation of the hypercolor group. In this way, we can recast them together as 
\begin{equation}
\psi=(U_L,D_L,\tilde{U}_L,\tilde{D}_L)^T~,
\end{equation}
which has a $SU(4)$ global (partially local) symmetry. Inside the $SU(4)$, 
the SM electroweak gauge group $SU(2)_W\times U(1)_Y$ is embedded in $SU(4)$ with generators given by
\begin{equation}\label{EWop}
SU(2)_W: \frac{1}{2}
\begin{pmatrix}
\sigma_a   &  0  \\
0   &  0  \\
\end{pmatrix},\quad
U(1)_Y: \frac{1}{2}
\begin{pmatrix}
0   &  0    \\
0   &  -\sigma_3^T     \\
\end{pmatrix}~.
\end{equation}
We can also find an $U(1)$ subgroup with the generator given by
\begin{equation}
\frac{1}{2\sqrt{2}}
\begin{pmatrix}
\mathbb{I}   &  0    \\
0   &  -\mathbb{I}   \\
\end{pmatrix},
\end{equation}
which is the same as hyperbaryon number $U(1)_{HB}$ by a factor of $N_{HC}/2\sqrt{2}$.

The hypercolor group becomes strongly coupled at the TeV scale, which forms a non-perturbative vacuum and breaks the global symmetry $SU(4)$ down to $Sp(4)$. Generically, there are two types of condensates it can form, which are
\begin{equation}
\Sigma_{EW}=
\begin{pmatrix}
i\sigma_2  &  0 \\
0   &  -i\sigma_2 \\
\end{pmatrix}
\quad\text{and}\quad
\Sigma_\slashed{EW}=
\begin{pmatrix}
0  &  \text{   }\mathbb{I}\text{   }  \\
-\mathbb{I}    &  0 \\
\end{pmatrix}.
\end{equation}
They both break the $SU(4)$ down to $Sp(4)$, but the first one preserves the two operators in eq.~\eqref{EWop} and the second one does not. In FCHMs, the condensate $\langle \psi\psi \rangle \propto \Sigma_{0}$ is chosen such that electroweak symmetry is preserved so the Higgs doublet will arise as pNGBs. Otherwise, it will become a Technicolor model, which is no longer favored.

Therefore, the $\Sigma_0$ should be the electroweak conserving vacuum $\Sigma_{EW}$. However, under this vacuum, the $U(1)_{HB}$ symmetry is broken. If the $U(1)_{HB}$ symmetry is global, we get a light scalar boson, which is the pNGB of broken $U(1)_{HB}$. Or, if the $U(1)_{HB}$ symmetry is local, a new massive $Z'$ vector boson will be generated, which is of our interest.

\subsection{Gauging the $U(1)'$ symmetry : $SM_3-HB$}

Gauging the hyperbaryon number is like gauging the SM baryon number, which is not a good idea. We can easily find that there is $SU(2)_W^2U(1)_{HB}$ anomaly. To cancel the quantum anomaly, we can make the SM fermions carry charges, too. We define a new $U(1)'$ gauge symmetry, under which SM fermions carry charge $Q_F$ for the species $F$, and hyperfermions carry the same charge as in $U(1)_{HB}$ but with an additional minus sign for the future convenience. To get rid of all the quantum anomalies, the charges of SM fermions need to satisfy
\begin{equation}
\displaystyle\sum_{F}^{{SM}} Q_{F}-N_{HC}\,Q_{HB}=0\quad\implies\quad
\displaystyle\sum_{F}^{{SM}} Q_{F}=N_{HC}\,Q_{HB}=1
\footnote{We assume that there is only one set of hyperfermions, i.e., a fourplet under $SU(2)_L\times SU(2)_R$, which are fundamental representations of hypercolor group $SP(N_{HC})$. The number can be other than 1 if there are more species of hyperfermion.}.
\end{equation}
Notice that the ${SM}$ counts the number of complete fourplet under $SU(2)_L\times SU(2)_R$, which includes 3 different colors of quarks and 1 lepton (sum up to 4) for each generation (total 3 generations)\footnote{Notice that the right-handed neutrinos are included in order to cancel the quantum anomalies. Therefore, the nature of neutrinos will also affect the properties of the $Z'$ boson, especially the $Z'\to \nu_R\bar{\nu}_R$ decay. In this work, we assume a Dirac-type neutrino for the following analysis.}. To simplify the model, we assume quarks and leptons carry a universal charge $Q_{SM}$\footnote{The charge can be non-universal without ruining the anomaly condition. For example, one can have a new $U(1)'$ as the linear combination of the original $U(1)'$ and an arbitrary SM anomaly-free $U(1)$, such as $U(1)_{B-L}$. In this case, the resulting $U(1)'$ will be non-universal but remain anomaly-free.}. Then, the anomaly cancellation requires $Q_{SM}=1/12$. However, to explain the neutral current B anomalies without violating the experimental constraints, we assume that only the third generation SM fermions $F_3$ carrying charge $Q_{SM_3}$. Then, the anomaly cancellation requires $Q_{SM_3}=1/4$ instead. Now we get an anomaly-free $U(1)'$ symmetry, which connects the SM sector and the hyperfermion sector. The form is similar to the anomaly-free $B-L$ in the SM, so we call this $U(1)'$ symmetry $SM_3-HB$. More discussions about the possible origin of this $U(1)'$ symmetry are included in Appendix \ref{sec:Origin}.

The interaction between the corresponding $Z'$ gauge boson and fermions is given by
\begin{align}\label{Zint0}
\mathcal{L}_{\text{int}}=g_{Z'}Z'_\mu\,(\,Q_{SM_3}\bar{F}_3\gamma^\mu F_3 - Q_{HB}\bar{\psi}_{}\gamma^\mu \psi_{}\,)=g_{Z'}Z'_\mu\left(\frac{1}{4}\,\bar{F}_3\gamma^\mu F_3 - \frac{1}{N_{HC}}\,\bar{\psi}_{}\gamma^\mu \psi_{}\right).
\end{align}
Now we get a new $Z'$ boson couples to the SM fermions and hyperfermions. On the hyperfermion side, their condensate will provide the $Z'$ mass, which is closed related to the Higgs and electroweak physics in the FCHM. On the SM side, since the third generation $F_3$ is based on the flavor basis, after the transformation, the $Z'$ boson will couple to all the SM fermions. Therefore, the $Z'$ boson can play a role in flavor physics and the neutral current B anomalies.

The $U(1)_{SM_3-HB}$ gauge symmetry, as the quantum number of the difference between the third generation SM fermion number and hyperbaryon number, is a new kind of $U(1)'$ symmetry people have not discussed before. However, this symmetry naturally arises when trying to build a fermionic UV theory for electroweak symmetry breaking. Furthermore, to have a light Higgs boson as a pNGB, this $U(1)'$ gauge symmetry must be broken at the TeV scale, which leads to a desired $Z'$ boson. Most important of all, this $Z'$ model opens the possibility to connect Higgs physics with flavor physics, which will be discussed in the following sections.

\subsection{The strength of $U(1)'$ gauge coupling}\label{sec:Strength}

For an $U(1)'$ gauge interaction, besides the quantum anomaly, we also need to take care of the Landau pole issue. The existence of a Landau pole implies that some new physics need to join in before we reach the scale, so it should not be too close to the TeV scale. The running of $g_{Z'}$ is calculated and expressed using $\alpha_{Z'}=g_{Z'}^2/4\pi$ as
\begin{align}
\alpha_{Z'}^{-1}(\mu)=\alpha_{Z'}^{-1}(\text{TeV})+\frac{b'}{2\pi}\,\text{ln}\left(\frac{\mu}{\text{TeV}}\right)
=\alpha_{Z'}^{-1}(\text{TeV})+0.37\,b'~\text{log}_{10} \left(\frac{\mu}{\text{TeV}}\right).
\end{align}
The value of $b'$ depends on the number of hypercolor $N_{HC}$ as 
\begin{align}
b'=-\frac{2}{3}\,\left[1+\frac{4}{N_{HC}}\right]~.
\end{align}
The coupling becomes non-perturbative when $\Lambda=10^n$ TeV, where 
\begin{align}
n=-\frac{\alpha_{Z'}^{-1}(\text{TeV})}{0.37\,b'}\approx\left(\frac{51\,N_{HC}}{4+N_{HC}}\right)\frac{1}{g_{Z'}^2 (\text{TeV})}~.
\end{align}
With the relation between the Landau pole and the value of $g_{Z'}$ at the TeV scale, we can calculate some bounds on $g_{Z'}$(TeV) for different constraints on the Landau pole. For example, taking $N_{HC}=2\,(b'=-2)$, if we want the Landau pole above the Planck scale, i.e., $\Lambda\gtrsim 10^{16}$ TeV, then at the TeV scale, $g_{Z'}\lesssim 1.0$ . If we relax the requirement to $\Lambda\gtrsim 10^3$ TeV, then $g_{Z'}$ can be up to $\sim 2.4$~. Taking $N_{HC}=4$ for the hypercolor group, where the $b'$ becomes smaller, the bounds on the $g_{Z'}$(TeV) in the two cases above become $\lesssim 1.3$ and $\lesssim 2.9$ respectively. These bounds will be applied in the analysis of section~\ref{sec:Combine}.

%\newpage

\section{The $SU(4)/Sp(4)$ fundamental composite Higgs model}\label{sec:Model}

We have already mentioned the $SU(4)/Sp(4)$ FCHM with its fermionic UV completion in the previous section as an example. Other choices include cosets $SU(5)/SO(5)$ \cite{Katz:2005au, Ferretti:2013kya, Vecchi:2013bja} and $SU(6)/SP(6)$ \cite{Katz:2005au, Cai:2018tet, Cheng:2020dum, Chung:2021fpc}. They both have a SM singlet pNGB, which corresponds to the broken hyperbaryon-like $U(1)$ symmetry. However, they also both have 14 pNGBs, which are too many for an economical model. Therefore, we will focus on the $SU(4)/Sp(4)$ FCHM, which have already been studied comprehensively in \cite{Galloway:2010bp, Cacciapaglia:2014uja, Cacciapaglia:2020kgq, Arbey:2015exa}. We will follow the previous studies to include the details of model building, the Higgs potential, and experimental constraints. The new phenomenology related to $Z'$ is left to section~\ref{sec:Z'}.

\subsection{Basics of $SU(4)/Sp(4)$}

To study the $SU(4)/Sp(4)$ coset, we can parametrize it by a nonlinear sigma model. Consider a sigma field $\Sigma$, which transforms as an anti-symmetric tensor representation $\mathbf{6}$ of $SU(4)$. The transformation can be expressed as $\Sigma \to g\,\Sigma \,g^T$ with $g\in SU(4)$. The scalar field $\Sigma$ has an anti-symmetric, electroweak conserving VEV $\langle \Sigma\rangle=  \Sigma_0$, where
\begin{equation}
\Sigma_0=\Sigma_{EW}=
\begin{pmatrix}
i\sigma_2  &  0 \\
0   &  -i\sigma_2 \\
\end{pmatrix}.
\end{equation}
The $\Sigma$ VEV breaks $SU(4)$ down to $Sp(4)$, producing 5 Nambu-Goldstone bosons.

The 15 $SU(4)$ generators can be divided into unbroken ones and broken ones with each type satisfying
\begin{equation}
\begin{cases}
\text{unbroken generators}   &T_a   : T_a\Sigma_0+\Sigma_0T_a^T=0~,\\
\text{broken generators}       &X_a   : X_a\Sigma_0-\Sigma_0X_a^T=0~.
\end{cases}
\end{equation}
The Nambu-Goldstone fields can be written as a matrix with the broken generator:
\begin{equation}
\xi(x)\equiv e^{\frac{i\pi_a(x)X_a}{2f}}~. 
\end{equation}
Under $SU(4)$, the $\xi$ field transforms as $\xi \to g\, \xi \,h^{\dagger}$ where $g \in SU(4)$ and $h \in Sp(4)$. The relation between $\xi$ and $\Sigma$ field is given by
\begin{equation}
\Sigma(x)= \xi\, \Sigma_0\,\xi^T=e^{\frac{i\pi_a(x)X_a}{f}}\Sigma_0~.
\end{equation}
The broken generators and the corresponding fields in the matrix can be organized as 
\begin{align}
i\pi_aX_a=&
\begin{pmatrix}
{ia}\,\mathbb{I} & \sqrt{2}\left(-H\tilde{H}\right) \\  
-\sqrt{2}\left(-H\tilde{H}\right)^\dagger & -ia\,\mathbb{I} \\  
\end{pmatrix}
\xrightarrow{\text{neutral component}}
\begin{pmatrix}
ia   &  0  &  0   &  h  \\
0   &  ia  &  -h   &  0  \\
0   &  h  &  -ia   &  0  \\
-h   &  0  &  0   &  -ia  \\
\end{pmatrix}~.
\label{Goldstone}
\end{align}
In this matrix, there are 5 independent fields. A Higgs (complex) doublet $H$ emerges as expected. There is one more singlet $a$, which corresponds to the broken hyperbaryon number, and it will become the longitudinal part of the $Z'$ boson. By these matrices, we can construct the Lagrangian for pNGB Higgs at low energy.

\subsection{The SM gauge sector}

The SM electroweak gauge group $SU(2)_W\times U(1)_Y$ is embedded in $SU(4)\times U(1)_X$ with generators given by
\begin{equation}
SU(2)_W: \frac{1}{2}
\begin{pmatrix}
\sigma_a   &  0  \\
0   &  0  \\
\end{pmatrix},\quad
U(1)_Y: \frac{1}{2}
\begin{pmatrix}
0   &  0  &  0   &  0  \\
0   &  0  &  0   &  0  \\
0   &  0  &  -1  &  0  \\
0   &  0  &  0   &  1  \\
\end{pmatrix} + X \mathbf{I}~.
\end{equation}
The extra $U(1)_X$ factor accounts for the different hypercharges of the SM fermion representations but is not relevant for the bosonic fields, of which we care more. These generators belong to $Sp(4)\times U(1)_X$ and are not broken by $\Sigma_0$. Using the $\Sigma$ field, the Lagrangian for kinetic terms of Higgs boson is given by
\begin{equation}
\mathcal{L}_h=\frac{f^2}{8}\text{tr}\left[(D_{\mu}\Sigma)(D^\mu \Sigma)^\dagger\right]+\cdots ,
\label{Lagrangian}
\end{equation}
where $D_{\mu}$ is the electroweak covariant derivative. Expanding this, we get
\begin{equation}
\mathcal{L}_h=\frac{1}{2}(\partial _\mu h)^2+\frac{f^2}{8}g_W^2 \,\text{sin}^2\left(\frac{h}{f}\right) \left[2W^+_\mu W^{-\mu}+\frac{Z_\mu Z^\mu}{\text{cos}^2\,\theta_W}\right],
\end{equation}
where $\theta_W$ is the usual weak mixing angle. When $h$ obtains a nonzero VEV $\langle h\rangle=V$, the $W$ and $Z$ bosons acquire masses of
\begin{equation}
m_W^2=\frac{f^2}{4}g_W^2 \,\text{sin}^2\left(\frac{V}{f}\right)=\frac{1}{4}g_W^2v^2,\quad m_Z^2=\frac{m_W^2}{\text{cos}^2\,\theta_W}
\end{equation}
where $v\equiv f\,\text{sin}(V/f)=246$ GeV.

The nonlinear behavior of the Higgs boson in the CHM is apparent from the dependence of trigonometric functions, which is parametrized by
\begin{equation}
\xi\equiv \frac{v^2}{f^2}= \sin^2\left(\frac{V}{f}\right)~.
\end{equation}
The Higgs boson couplings to SM fields are modified by the nonlinear effect due to the pNGB nature of the Higgs boson. For example, the deviation of the Higgs coupling to vector bosons is parameterized by
\begin{equation}
\kappa_V\equiv \frac{g_{hVV}}{g^{SM}_{hVV}}=
\text{cos}\left(\frac{V}{f}\right)=\sqrt{1-\xi}\approx 1-\frac{\xi}{2}~,
\end{equation}
which provides the most important test of CHMs.

%\newpage
\subsection{The $U(1)'$ gauge symmetry}\label{sec:U(1)}

When $SU(4)$ global symmetry is broken down to $Sp(4)$ by the $\Sigma_0$, the $U(1)_{HB}$ subgroup as well as $U(1)'$ gauge symmetry is also broken down, which results in a massive gauge boson $Z'$. The mass of $Z'$ gauge boson is generated through the term
\begin{equation}\label{hZ'}
\mathcal{L}_{Z'}=2f^2\,Q_{HB}^2\,g_{Z'}^2 \,\text{cos}^2\left(\frac{h}{f}\right) {Z'_\mu Z'^\mu},
\end{equation}
where the cosine term comes from the vacuum misalignment due to the nonzero VEV $\langle h\rangle=V$. The $Z'$ mass is given by
\begin{equation}\label{Zmass}
M_{Z'}=2\,Q_{HB}\,g_{Z'}f\,\text{cos}\left(\frac{V}{f}\right)=\frac{2}{N_{HC}}g_{Z'}f\,\text{cos}\left(\frac{V}{f}\right),
\end{equation}
where $Q_{HB}=1/N_{HC}$ is substituted in. For future convenience, we define the $Z'$ scale 
\begin{equation}
f'\equiv \frac{M_{Z'}}{g_{Z'}}=\frac{2}{N_{HC}}f\,\text{cos}\left(\frac{V}{f}\right)\approx \frac{2}{N_{HC}}f,
\end{equation}
which is relevant in the study of the $Z'$ phenomenology in section~\ref{sec:Z'}. The relation depends on the number of hypercolor $N_{HC}$. If $N_{HC}=2$, the $Z'$ scale $f'$ is close to the symmetry breaking scale $f$ up to a cosine factor which is close to one. Another case we will discuss is $N_{HC}=4$, where $f'\approx f/2$ and it corresponds to $SP(4)$ hypercolor group.\footnote{The $N_{HC}=2$ case has been assumed and studied in the previous work \cite{Chung:2021ekz}. In this work, we do the calculation based on a general $N_{HC}$ and compare the difference between the two cases we focus on.}

The Higgs coupling to the $Z'$ boson can also be derived from eq. \eqref{hZ'} as
\begin{equation}\label{hZ'Z'}
g_{hZ'Z'} =-8\,Q_{HB}^2\,g_{Z'}^2v\sqrt{1-\xi},\quad g_{hhZ'Z'} =-4\,Q_{HB}^2\,g_{Z'}^2({1-2\xi}),
\end{equation}
where the minus sign plays an important role when calculating the Higgs potential.

%\newpage
\subsection{The Yukawa sector}

There are many mechanisms to generate Yukawa couplings in the composite Higgs models. No matter which, new interactions are required to connect the SM fermion sector with the strong sector. The widely used scenario is through the partial compositeness mechanism~\cite{Kaplan:1991dc}, where elementary SM fermions mix with composite operators of the same SM quantum numbers from the strong dynamics. The terms $\mathcal{L}_{\text{mix}}=\lambda_{L}\bar{q}_{L}{O}_{R}+\lambda_{R}\bar{q}_{R}{O}_{L}$, where $q_L$, $q_R$ are elementary SM fermions and ${O}_L$, ${O}_R$ are composite operators of some representations of the global symmetry $G$. With these interactions, the observed SM fermions will be mixtures of elementary fermions and composite resonances. The SM fermions then couple to the Higgs field through the portion of the strong sector. However, they usually require light composite fermionic top partners, which are challenging both theoretically and experimentally.

On the theory side, additional mechanisms are required to bring the top partner mass well below the generic scale $\Lambda\sim4\pi f$. Also, the constructions of the fermionic UV completion require many extensions in the strong sector~\cite{Barnard:2013zea, Ferretti:2013kya}. On the experiment side, the current bound on the top partner mass has reached $\sim 1.2$~TeV~\cite{Sirunyan:2018omb, Aaboud:2018pii}, which is also larger than the value preferred by the naturalness principle.

In this work, we assume the connection between the SM sector and the strong sector is through bilinear coupling, which can be realized in models with extended hypercolor \cite{Cacciapaglia:2015yra}. In this scenario, the top Yukawa coupling is generated by the 4-fermion operator
\begin{equation}\label{Ltop}
\mathcal{L}_{t}=\frac{1}{\Lambda_{t}^2}\left(Q_L\bar{t}_R\right)^\dagger_a\left(\psi^TP^a\psi\right),
\end{equation}
where $\Lambda_{t}$ is some dynamical scale responsible for the top Yukawa, and $a$ is an $SU(2)_W$ index. The projectors are given by
\begin{equation}
P^1: \frac{1}{2}
\begin{pmatrix}
0   &  0  &  1   &  0  \\
0   &  0  &  0   &  0  \\
-1   &  0  &  0  &  0  \\
0   &  0  &  0   &  0  \\
\end{pmatrix},\quad
P^2: \frac{1}{2}
\begin{pmatrix}
0   &  0  &  0   &  0  \\
0   &  0  &  1   &  0  \\
0   &  -1  &  0  &  0  \\
0   &  0  &  0   &  0  \\
\end{pmatrix},
\end{equation}
which picks up the Higgs component inside the $\psi^T\psi$. The resulting top Yukawa coupling is roughly 
\begin{equation}\label{yt}
y_t\sim \frac{1}{\Lambda_{t}^2}\left(4\pi f^2\right).
\end{equation}
The value of $y_t$ can be derived directly if the UV completion of these operators is given. However, the UV model is beyond the scope of the present work and is left to a future study. In this study, we take a bottom-up approach and assume $y_t\sim m_t/\sqrt{2}v$. Then, the Lagrangian of top Yukawa is given by
\begin{equation}
\mathcal{L}_{t}= -\frac{1}{\sqrt{2}}y_t\bar{t}_Lt_R\,f \text{sin}\left(\frac{h}{f}\right)
=-m_t\bar{t}_Lt_R-\kappa_t\frac{m_t}{v}h\bar{t}_Lt_R+\cdots~,
\end{equation}
where
\begin{equation}
\kappa_t\equiv \frac{g_{h\bar{t}t}}{g^{SM}_{h\bar{t}t}}=
\text{cos}\left(\frac{V}{f}\right)=\sqrt{1-\xi}\approx 1-\frac{\xi}{2}~.
\end{equation}
Again, we see the deviation from the SM prediction in the Higgs coupling due to its pNGB nature. We assume all the SM fermions couple to the Higgs and get their masses through a similar mechanism. Therefore, we have $\kappa_F=\kappa_t=\sqrt{1-\xi}$ for all the SM fermions, which is also the same as $\kappa_V$ in the gauge sector.

%\newpage
\subsection{The loop-induced Higgs potential}

The Higgs potential in $SU(4)/Sp(4)$ CHMs is generated from the $SU(4)$ symmetry breaking sources, mainly from the interactions with other fields. In this model, the main contributions come from the gauge coupling and top Yukawa coupling. The CHM Higgs potential can be parameterized by 
\begin{equation}\label{Vh}
V(h)={\alpha} f^2\text{sin}^2\left(\frac{h}{f}\right)+{\beta}f^4\,\text{sin}^4\left(\frac{h}{f}\right)~.
\end{equation}
where ${\alpha}$ has mass dimension two and ${\beta}$ is dimensionless, similar to the coefficients of the quadratic and quartic term in SM Higgs potential. Together with the scale $f$, the three parameters need to satisfy
\begin{equation}\label{mass}
v=\sqrt{\frac{{-\alpha}}{2{\beta}}}, \qquad m_h^2=8{\beta} {v^2}(1-\frac{v^2}{f^2})~.
\end{equation}
Since we already know the observed Higgs is SM-like, the scale $f$ should be much larger than $v$. Therefore, the required coefficient is approximately
\begin{equation}\label{eq:2}
\alpha\simeq-\left(63\text{ GeV}\right)^2,\qquad
\beta\simeq 0.033~.
\end{equation}
Knowing the required values in the CHM Higgs potential, we can then discuss the contribution from each interaction and the fine-tuning issue.

\subsubsection{The gauge contribution}

The gauge interactions explicitly break the $SU(4)$ global symmetry, so they contribute to the potential of the Higgs field. The gauge bosons couple to pNGBs through the mixing with composite resonances as
\begin{equation}
\mathcal{L}_{\text{mix}}=g_WW_{\mu,a}J^{\mu,a}_W+g_YB_{\mu}J^{\mu}_Y+g_{Z'}Z'_{\mu}J^{\mu}_{Z'}~,
\end{equation}
where the $J_W$, $J_Y$, and $J_{Z'}$ belong to the composite operators in an adjoint representation $\mathbf{15}$ of $SU(4)$. After the symmetry breaking, the composite operators are decomposed into $\mathbf{10}$ and $\mathbf{5}$ of $Sp(4)$. The masses of composite resonances of different $Sp(4)$ representations are different in general, which will generate a potential for pNGBs at $\mathcal{O}(g^2)$.

Starting from the $SU(2)_W$, it only breaks the global symmetry partially and generates mass terms for the Higgs doublet:
\begin{align}
SU(2)_W: ~c_w\frac{3}{16\pi^2}\frac{3g_W^2}{4}M_\rho^2 = c_w\frac{9}{64\pi^2}g_W^2g_\rho^2f^2,
\end{align}
where $M_\rho\sim g_\rho f $ is the mass of the vector resonances $\rho$ which act as the gauge boson partners, and $c_w$ is a $\mathcal{O}(1)$ constant. Similarly, for $U(1)_Y$ and $U(1)'$, the interactions also break the global symmetry partially. They generate mass terms for $H$:
\begin{align}
U(1)_Y:& ~c_y\frac{3}{16\pi^2}\frac{g_Y^2}{4}M_{\rho}^2= c_y\frac{3}{64\pi^2}g_Y^2g_{\rho}^2f^2,\\
U(1)':& ~-c'\frac{3}{16\pi^2}4\,Q_{HB}^2\,g_{Z'}^2M_{\rho}^2= 
-c'\frac{3}{4\pi^2}Q_{HB}^2\,g_{Z'}^2g_{\rho}^2f^2,
\end{align}
where $c_y$ is also an $\mathcal{O}(1)$ constant but $c'$ could be smaller if $Z'$ boson is heavy.

Combining these three contributions, we get the mass terms of the pNGBs from the gauge contributions at the leading order as
\begin{align}\label{pNGBmass}
\alpha_g=\left({\frac{3}{4}g_W^2c_w+\frac{1}{4}g_Y^2c_y-4\,Q_{HB}^2\,g_{Z'}^2c'}\right)\frac{3}{16\pi^2} g_{\rho}^2f^2~.
\end{align}
Notice that the contribution from the $U(1)'$ is opposite to the SM gauge interaction, which can also be seen from the minus sign in eq~\eqref{hZ'Z'}. The reason it gives an opposite contribution can be understood as that the $U(1)'$ gauge symmetry wants to stay unbroken and thus prefers the other vacuum, i.e. the electroweak breaking vacuum $\Sigma_\slashed{EW}$. The negative contribution will make the original vacuum unstable and push it to the other vacuum that preserves the $U(1)'$ gauge symmetry.

\subsubsection{The top contribution}

Next, the other important contribution comes from the top quark loop. The top-loop effects on the pNGB potential can be traced with the spurions $P^a$ in eq. \eqref{Ltop}. The leading order contribution from the top quark loop is
\begin{equation}\label{H2_mass}
\sim -\frac{N_c}{16 \pi^2} y_t^2M_t^2 f^2 \left|{(P)^a}_{ij}\Sigma^{ij}\right|^2+\frac{N_c}{16 \pi^2} y_t^4 f^4 \left|{(P)^a}_{ij}\Sigma^{ij}\right|^4~,
\end{equation}
and thus the contribution to the Higgs quadratic and quartic term is given by 
\begin{align}\label{eq:alpha}
\alpha_t=-c_t\frac{N_c}{16\pi^2} y_t^2 M_t^2,\qquad
\beta_t=\frac{N_c}{16\pi^2}y_t^4~,
\end{align}
where $M_t$ is some mass scale related to the top loop and $c_t$ is a $\mathcal{O}(1)$ constant. Usually, in the model with top partners such as CHMs with partial compositeness, $M_t$ should be the mass of top partners which cancels out the divergent top loop. However, in our case where the top Yukawa is generated through the 4-fermion operator eq.~\eqref{Ltop}, $M_t$ would be the mass of the new degree of freedom that generates this operator, above which the top Yukawa no longer exists. Also, the mass scale $M_t$ here differs from the dynamical scale $\Lambda_{t}$ in eq.~\eqref{Ltop} by a factor of coupling $g_t$, i.e. $M_t=g_t \Lambda_{t}$.

\subsubsection{Fine-tuning in the Higgs potential}\label{sec:Tuning}

The observed light Higgs implies a small quartic, which can be generated from the loop-induced potential. The $\beta_t$ from the top quark loop already gives a comparable value. The problem of the Higgs potential in CHMs is that the loop-induced $\alpha$ is too large as
\begin{align}\label{quadratic}
\alpha=\alpha_g+\alpha_t=\left({\frac{3}{4}g_W^2c_w+\frac{1}{4}g_Y^2c_y-4\,Q_{HB}^2\,g_{Z'}^2c'}\right)\frac{3}{16\pi^2} M_{\rho}^2-c_t\frac{N_c}{16\pi^2} y_t^2 M_t^2.
\end{align}
The required $\alpha$ is $\sim -\left(63\text{ GeV}\right)^2$, which implies the generic $M_t\lesssim 500$ GeV. However, none of the new physics has been observed yet. Some cancellations among coefficients are required to raise $M_t$, i.e., fine-tuning. The fine-tuning issue is beyond the scope of this study. In this study, we only want to make a few claims to bring up the disfavored region in the parameter space.

Without specifying the new physics and exotic states, we can see the potential is loop-induced. Therefore, the generic scale for $f$ should be $\sim 4\pi \sqrt{-\alpha}\sim 800$ GeV. However, if we take $M_t\sim f$, it already needs some $30\%$ fine-tuning due to the large top Yukawa coupling and $N_c=3$. If we enhance the value by a factor of two to $f\sim1.6$ TeV, the required tuning becomes $\lesssim 10\%$, which is considered unnatural. Therefore, in this paper, we will take the value $f=1.6$ TeV as the bound above which the parameter space is theoretically disfavored in section~\ref{sec:Combine}.

%\newpage
\subsection{Exprimental constraints}\label{sec:HiggsEW}

There are many directions when talking about the experimental constraints of a composite Higgs model. The direct searches for a new particle are, of course, the most intriguing ones. Especially, there are tons of composite resonances around compositeness scale $\Lambda$ in a CHM to be discovered. However, the first and the most promising target should be the light state, like pNGBs. Higgs boson as one of them has already been discovered. The next one should be the $Z'$ boson. A detailed study of the $Z'$ phenomenology, including direct and indirect searches, will be presented in section~\ref{sec:Z'}. Besides direct searches, precision tests also provide strong bounds on the CHMs due to the modification of the Higgs sector. In this section, we will focus on the bounds from Higgs coupling measurement and electroweak precision tests.

\subsubsection{Higgs coupling measurement}

The key feature of composite Higgs models is the nonlinear nature of pNGB, which can be directly detected through the Higgs coupling measurement. In our model, the modification of the Higgs couplings is
\begin{equation}
\kappa=\kappa_V=\kappa_F=
\text{cos}\left(\frac{V}{f}\right)=\sqrt{1-\xi}\approx 1-\frac{\xi}{2}~.
\end{equation}

The current best-fit values of $\kappa_V$ and $\kappa_F$ from ATLAS \cite{ATLAS:2020qdt} with an integrated luminosity of 139 fb$^{-1}$ are
\begin{align}
&\kappa_V=1.03\pm 0.03~,\\
&\kappa_F=0.97\pm 0.07~,
\end{align}
with a 50\% correlation between the two quantities. The central values for both quantities are close to the SM value, but the uncertainties are still large. Considering all the decay channels, the bound on the value of $\xi$ and the scale $f$ at 95\% CL level is given by \cite{Khosa:2021wsu}
\begin{equation}\label{Higgsbound}
\xi \leq 0.1, \qquad f \geq 780 ~\text{GeV .}
\end{equation}
In the future, the uncertainties in $\kappa_V$ and $\kappa_F$ can be improved to 1\% and 3\% respectively during the HL-LHC era~\cite{deBlas:2019rxi}. Assuming the central values are the same, it can probably rule out all the CHMs at 99\% CL due to the current large $\kappa_V$. The next generation Higgs factories, such as ILC, CEPC, and FCCee, will have great sensitivities to the $hZZ$ coupling and will measure $\kappa_V$ with a precision $\approx$ 0.3\%. Even with the central value at $(1,1)$, it can test the symmetry breaking scale $f$ up to several TeVs and hence cover the entire natural parameter space for CHMs.

\subsubsection{Oblique Parameters}

The electroweak oblique corrections provide important tests of new physics related to the electroweak sector around the TeV scale. They are usually expressed in terms of $S$, $T$, and $U$ parameters~\cite{Peskin:1990zt,Peskin:1991sw}. For heavy new physics, $U$ is typically small as it is suppressed by an additional factor $M_{\rm new}^2/m_Z^2$. Therefore, we will only focus on the contributions to the $S$ and $T$ parameters in this section.

\subsubsection*{The modification of Higgs couplings} 

In CHMs, there is a contribution to both $S$ and $T$ from the nonlinear Higgs dynamics due to the modifications of the Higgs couplings, which result in an incomplete cancellation of the electroweak loops~\cite{Barbieri:2007bh,Grojean:2013qca}. This contribution is proportional to $\xi$ and depends logarithmically on $M_\rho/m_h$ as 
\begin{align}
\Delta S=\frac{1}{12\pi}\text{ ln}\left(\frac{M_\rho^2}{m_h^2}\right)\cdot\xi,\quad
\Delta T=-\frac{3}{16\pi c_w^2}\text{ ln}\left(\frac{M_\rho^2}{m_h^2}\right)\cdot\xi
\end{align}
Taking $M_\rho=5$~TeV, we get
\begin{align}\label{ST1}
\Delta S\sim 0.20 ~\xi,\quad \Delta T\sim -0.57 ~\xi~.
\end{align}

\subsubsection*{The contribution from hyperfermions} 

In FCHMs, the hyperfermions, which carry electroweak quantum numbers, also contribute to the $S$ parameter. We can estimate a rough value by calculating the one-loop contribution from the hyperfermions with heavy constituent mass terms. In this case, each $SU(2)_W$ doublet contributes $\sim 1/6\pi$ and gives
\begin{align}\label{ST2}
\Delta S\sim \frac{1}{6\pi}N_{HC}~\xi\sim 0.11~\xi \quad(if~N_{HC}=2)~.
\end{align}

So far, these contributions are common in FCHMs and are under control. If we only consider them, the overall contribution to the oblique parameters is given by
\begin{align}\label{ST}
\Delta S\sim 0.31 ~\xi,\quad \Delta T\sim -0.57 ~\xi~.
\end{align}
The positive $\Delta S$ and negative $\Delta T$ point to the direction with the most stringent constraint. However, there are also additional sources when considering composite resonances in the model, which can change the direction.

\subsubsection*{The contribution from additional sources} 

The $S$ parameter also receives a contribution from the mixing between the SM gauge bosons and the composite vector resonances. It is estimated to be~\cite{Agashe:2003zs,Agashe:2005dk,Giudice:2007fh}
\begin{align}
\Delta S \sim c_S~4\pi \frac{v^2}{M_\rho^2}  
\sim c_S~\left(\frac{4\pi }{g_\rho^2}\right)~\xi~,
\end{align}
where $c_S$ is an $\mathcal{O}(1)$ factor. If we take $c_S=1$ and $g_\rho\sim 2\pi$, it gives the contribution $\Delta S\sim 0.32~\xi$, which turns out to be the dominant contribution on the $S$ parameter. However, another positive contribution on $\Delta S$ does not help. To bring the deviation back to the safe region, an additional positive $\Delta T$ is required.

The $T$ parameter parametrizes the amount of custodial $SU(2)$ breaking, which receives a contribution from the mixing of the hypercharge gauge boson and vector resonances. However, it is small due to the custodial symmetry, which makes the tree-level contribution vanish and the loop contribution negligible. In the CHMs with partial compositeness, there are additional contributions to $\Delta T$ from composite fermion resonances. The dominant contribution is through the top partners, which is closely related to the origin of the top Yukawa coupling. However, the exact value of $\Delta T$ depends on the complete spectrum of composite fermion resonances, which is model-dependent. A comprehensive analysis has been studied in \cite{Frandsen:2022xsz}.

The current global fit of the oblique parameters is given by~\cite{ParticleDataGroup:2020ssz}
\begin{equation}
S=-0.01 \pm 0.10, \quad T=0.03 \pm 0.12, \quad U= 0.02\pm 0.11~.
\end{equation}
If one fixes $U=0$, then $S$ and $T$ constraints improve to
\begin{equation}
S=0.00 \pm 0.07, \quad T=0.05 \pm 0.06~,
\end{equation}
with a strong positive correlation (92\%) between them. The latest $M_W$ measurement by the CDF collaboration \cite{CDF:2022hxs}, however, shifts the best fit values upward as \cite{Lu:2022bgw, Strumia:2022qkt, deBlas:2022hdk, Asadi:2022xiy, Bagnaschi:2022whn}
\begin{equation}
S=0.06 \pm 0.08, \quad T=0.15 \pm 0.06~,
\end{equation}
which imply a larger custodial symmetry breaking.

The resulting constraints on $SU(4)/Sp(4)$ Fundamental Composite Higgs Models with partial compositeness, including different kinds of top-partner spectrums, have been studied in \cite{Frandsen:2022xsz}. It was shown that there is a good agreement between the current electroweak precision measurements and the $SU(4)/Sp(4)$ FCHMs even with a small $\xi \sim 0.1$, which means the bound from the EWPTs is actually weaker than the bound from the Higgs coupling measurement. Even without partial compositeness and top partners, if we want to explain the top-bottom mass hierarchy, some custodial symmetry violating mechanism must be introduced. Otherwise, it will not be able to distinguish $t_R$ and $b_R$. Therefore, we can also expect some additional contributions to $\Delta T$ in other kinds of CHMs.

In summary, due to the corrections from numerous sources of various UV-completion, the bound from electroweak precision tests is hard to be determined and can be quite weak. Therefore, we will only consider the bound from the Higgs coupling measurement given in eq.~\eqref{Higgsbound} in the following discussion.

\section{$Z'$ phenomenology}\label{sec:Z'}

The previous discussions on the FCHM and the resulting Higgs and electroweak physics give us the constraints on the symmetry breaking scale $f$. Next, we move on to explore the most relevant new particle - the $Z'$ boson. Especially, we will use it to explain the observed neutral current B anomalies. Following the analysis in \cite{Chung:2021ekz}, we will analyze the constraints on its scale $f'$, which come from flavor physics, and its mass $M_{Z'}$, which come from LHC direct searches in this section.

\subsection{$Z'$ couplings to SM fermions}\label{sec:Coupling}

To discuss the $Z'$ phenomenology, we need to first transform the $Z'$ interaction terms in eq.~\eqref{Zint0} to the physical basis. First, we rewrite it to include all generations and chiralities in the flavor basis as
\begin{align}\label{Zint1}
\mathcal{L}_{\text{int}}=g_{Z'}Z'_\mu\,(\,\bar{F}_L^f\gamma^\mu  Q_{F_{L}}^fF_L^f+\bar{F}_R^f\gamma^\mu  Q_{F_{R}}^fF_R^f\,),
\end{align}
where $F=(F_1,F_2,F_3)$ and $f$ denotes the flavor basis. The charge matrices are given by
\begin{equation}\label{Zint2}
Q_{F_{L/R}}^f=Q_{SM_3}
\begin{pmatrix}
0   &  0  &  0    \\
0   &  0  &  0    \\
0   &  0  &  1    \\
\end{pmatrix}=\frac{1}{4}
\begin{pmatrix}
0   &  0  &  0    \\
0   &  0  &  0    \\
0   &  0  &  1    \\
\end{pmatrix}.
\end{equation}
Next, we need to transform them to the mass basis $F^m_{L/R}$ through the mixing matrices $U_{F_{L/R}}$, which satisfy $F^f_{L/R}= U_{F_{L/R}} F^m_{L/R}$. After the transformation, we get
\begin{align}
\mathcal{L}_{\text{int}}=g_{Z'}Z'_\mu\,(\,\bar{F}_L^m\gamma^\mu Q_{F_{L}}^mF_L^m+\bar{F}_R^m\gamma^\mu Q_{F_{R}}^mF_R^m\,),
\end{align}
where the charge matrices become
\begin{equation}
Q_{F_{L/R}}^m=U_{F_{L/R}}^\dagger Q_{F_{L/R}}^fU_{F_{L/R}}.
\end{equation}

To determine the magnitude of each interaction, we need to know all the $U_{F_{L/R}}$. However, The only information about these unitary transformation matrices is given by the CKM matrix for quarks and the PMNS matrix for leptons, which are given by
\begin{equation}\label{CKM}
V_{CKM}= U_{u_L}^\dagger U_{d_L}\quad\text{and}\quad
V_{PMNS}= U_{e_L}^\dagger U_{\nu_L}.
\end{equation}
These relations only tell us about the left-handed part without any right-handed information. Even with these two constraints, we only know the difference between two unitary transformations, but none for the individual one. Therefore, we need to make some assumptions about these matrices to go further.

To simplify the analysis, we assume all the right-handed matrices $U_{F_{R}}$ are identity matrices. Therefore, for the right-handed fermions, only the third generation participates in the $Z'$ interaction, which means there is no FCNC. The couplings are universal for all the right-handed third generation SM fermions with value $g_{Z'}/4$.

For the left-handed side, due to the observation of nontrivial $V_{CKM}$ and $V_{PMNS}$, there must be nonzero rotations for $U_{F_{L}}$. A general rotation matrix should include three angles regardless of phases. However, to minimize the required parameters, we only specify the rotation between the second and third generations. Furthermore, since we want to address the neutral current B anomalies and related measurements, we will focus on the rotation matrices of the down-type quarks, $U_{d_L}$, and charged leptons. $U_{e_L}$. The specified matrices are given by
\begin{equation}
U_{d_L}=
\begin{pmatrix}
1   &0   &0    \\
0   &\text{cos}~\theta_d   &  \text{sin}~\theta_d \\
0   &-\text{sin}~\theta_d  &  \text{cos}~\theta_d \\
\end{pmatrix},\quad
U_{e_L}=
\begin{pmatrix}
1   &0   &0    \\
0   &\text{cos}~\theta_e   &  \text{sin}~\theta_e \\
0   &-\text{sin}~\theta_e  &  \text{cos}~\theta_e \\
\end{pmatrix}~.
\end{equation}
Notice that, although they look similar, the magnitudes we expect for the two angles could be quite different. For $\theta_d$, we expect it to be CKM-like. That is sin$\,\theta_d\sim \mathcal{O}(0.01)$. For $\theta_e$, it could be around sin$\,\theta_e\sim \mathcal{O}(0.1)$ or larger. With the assumption, the other two matrices are also fixed by the relation in eq. \eqref{CKM} as
\begin{equation}
U_{u_L}= U_{d_L} V_{CKM}^\dagger\quad\text{and}\quad 
U_{\nu_L}= U_{e_L} V_{PMNS}.
\end{equation} 
We can find that $U_{u_L}(U_{\nu_L})$ is similar to $V_{CKM}(V_{PMNS})$ up to a $\theta_d\,(\theta_e)$ rotation.

With the specified matrices, the resulting charge matrices are given by
\begin{equation}
Q_{d_{L}}=\frac{1}{4}
\begin{pmatrix}
0   &  0  &  0    \\
0   &   \text{sin}^2\,\theta_d  &  -\, \text{sin}\,\theta_d\,\text{cos}\,\theta_d    \\
0   &  -\, \text{sin}\,\theta_d\,\text{cos}\,\theta_d  &   \text{cos}^2\,\theta_d    \\
\end{pmatrix},\quad
Q_{e_{L}}=\frac{1}{4}
\begin{pmatrix}
0   &  0  &  0    \\
0   &   \text{sin}^2\,\theta_e  &  -\, \text{sin}\,\theta_e\,\text{cos}\,\theta_e    \\
0   &  -\, \text{sin}\,\theta_e\,\text{cos}\,\theta_e  &   \text{cos}^2\,\theta_e    \\
\end{pmatrix}~.
\end{equation}
Then, we can write down all the couplings with the $Z'$ boson for the left-handed fermions. To study the B anomalies, two of them, $Z'sb$ and $Z'{\mu\mu}$ are especially important so we define the corresponding couplings as $g_{sb}$ and $g_{\mu\mu}$. Moreover, we can extract the charge and mixing part of the couplings and rewrite the couplings as
\begin{align}
g_{sb}= -\frac{1}{4}\,g_{Z'}\epsilon_{sb} &\quad\text{where}\quad
\epsilon_{sb}=\, \text{sin}\,\theta_d\,\text{cos}\,\theta_d~,\\
g_{\mu\mu}= \frac{1}{4}\,g_{Z'}\epsilon_{\mu\mu}
&\quad\text{where}\quad\epsilon_{\mu\mu}=\text{sin}^2\,\theta_e~.
\end{align}
We will see later that the constraints from flavor physics will be put on the three key parameters: the scale $f'$, the mixings $\epsilon_{sb}$, and $\epsilon_{\mu\mu}$.

%\newpage

\subsection{Flavor Phenomenology}\label{sec:Flavor}

With the specified mixing matrices, we can then explore the consequence in flavor physics, especially the parameter space allowed to explain the neutral current B anomalies. Also, the constraints from other low energy experiments, including neutral meson mixings and lepton flavor violating decays, are shown in this section.

\subsubsection{Neutral Current B Anomalies}

To explain the observed neutral current B anomalies, an additional contribution on $b\to s\mu\mu$ is required. Based on the assumption we made, after integrating out the $Z'$ boson, we can get the operator
\begin{equation}
\Delta\mathcal{L}_{\text{NCBA}}=\frac{4G_F}{\sqrt{2}}V_{tb}V_{ts}^*\frac{e^2}{16\pi^2}\,C_{LL}(\bar{s}_L\gamma_\mu b_L)(\bar{\mu}_L\gamma^\mu\mu_L)
\end{equation}
in the low energy effective Lagrangian with coefficient
\begin{equation}
C_{LL}
=\frac{g_{sb}g_{\mu\mu}}{M_{Z'}^2}~(36~\text{TeV})^2
=-\frac{\epsilon_{sb}\epsilon_{\mu\mu}}{f'^2}~(9~\text{TeV})^2.
\label{Banomaly}
\end{equation}

The global fit value for the Wilson coefficient $C_{LL}$, considering only the clean observables \cite{Altmannshofer:2021qrr}, is
\begin{equation}\label{fitting}
C_{LL}=-0.70\pm 0.16~,
\end{equation}
which requires
\begin{subequations}
\label{bsmumu}
\begin{equation}
\frac{\epsilon_{sb}\epsilon_{\mu\mu}}{f'^2}=\frac{1}{(11~\text{TeV})^2}
%\left(\frac{C_{LL}}{-0.70}\right)
\implies f'= \sqrt{\epsilon_{sb}\epsilon_{\mu\mu}%\left(\frac{-0.70}{C_{LL}}\right)
}~(11~\text{TeV}).
\end{equation}
Besides, we also consider the value $C_{LL}=-0.38$, which is $2\sigma$ above the central value. This value represents a smaller contribution from new physics but is of experimental and theoretical interest as we will see in the following sections. It requires
\begin{equation}
\frac{\epsilon_{sb}\epsilon_{\mu\mu}}{f'^2}=\frac{1}{(14~\text{TeV})^2}
%\left(\frac{C_{LL}}{-0.70}\right)
\implies f'= \sqrt{\epsilon_{sb}\epsilon_{\mu\mu}%\left(\frac{-0.70}{C_{LL}}\right)
}~(14~\text{TeV}).
\end{equation}
\end{subequations}
The two different values of $C_{LL}$ will lead to very different phenomenology, especially in the direct search strategies, which will be discussed in the following sections.

The generic scale for both cases shows $f' \gtrsim 10$ TeV. However, as we mention in the last section, the value $\epsilon_{sb}\sim\mathcal{O}(0.01)$ can bring it down to the TeV scale, where we expect the solution to the hierarchy problem.

\subsubsection{Neutral Meson Mixing}

Next, we discuss the constraints from other measurements. Strong constraints on the $Z'$ solution come from neutral meson mixings. In our specified mixing matrices, the mixings between the first generation down-type quark and other generations are not presented, so there are no additional contributions on $K^0-\bar{K}^0$ and $B_d-\bar{B}_d$ mixings. In the up-type quark sector, additional contribution on $D^0-\bar{D}^0$ mixing receives CKM suppression, so the constraint is also relaxed. Therefore, the $B_s-\bar{B}_s$ mixing, which contains the second and third generation down-type quarks, turns out to give the strongest bound on the $Z'$ coupling. The contribution from the $Z'$ exchange comes from the operator
\begin{equation}
\Delta\mathcal{L}_{B_s}=-\frac{1}{2}\frac{g_{sb}^2}{M_{Z'}^2}(\bar{s}_L\gamma_\mu b_L)(\bar{s}_L\gamma_\mu b_L).
\end{equation}
Following the calculation in \cite{DiLuzio:2017fdq}, assuming $M_{Z'}$ is around the TeV scale, we can derive the deviation on the mass difference of neutral $B_s$ mesons as
\begin{equation}
C_{B_s}\equiv \frac{\Delta M_s}{\Delta M_s^{SM}}
\approx 1+5576\,\left(\frac{g_{sb}}{M_{Z'}}\right)^2,
\end{equation}

The measurement of mixing parameter \cite{HFLAV:2019otj} compared with SM prediction by sum rule calculations \cite{King:2019lal} gives a strong upper bound at 95\% C.L. as \cite{Allanach:2019mfl}
\begin{equation}\label{Bsmixing}
\left|\frac{g_{sb}}{M_{Z'}}\right| \leq \frac{1}{194~\text{TeV}}\implies
\epsilon_{sb} \leq \frac{f'}{ 48.5~\text{TeV}}~.
\end{equation}
Next, we can combine it with the requirement from eq. \eqref{bsmumu}, which allows us to transfer the upper bound on $\epsilon_{sb}$ to the lower bound on $\epsilon_{\mu\mu}$ as
\begin{equation}\label{minmumu}
\epsilon_{\mu\mu} \geq \frac{f'}{ 2.3~(4.3)~\text{TeV}}~,
\end{equation}
where the value corresponds to $C_{LL}=-0.70\,(-0.38)$ respectively.

The transformation means: for the $b\to s\mu\mu$ process, the $bs$ side, which is constrained by the $B_s-\bar{B}_s$ mixing measurement, is extremely suppressed. Therefore, the $\mu\mu$ side needs to be large enough to generate the observed neutral current B anomalies. Therefore, with smaller $C_{LL}$, the required $ \epsilon_{\mu\mu}$ is also smaller.

Moreover, because there is a maximum value for $\epsilon_{\mu\mu}=$ sin$^2\,\theta_e\leq 1$, it implies an upper bound $2.3\,(4.3)$ TeV for the scale $f'$, which is consistent with the scale we expected. Also, we can combine eq. \eqref{Bsmixing} and eq. \eqref{minmumu}, which gives us the ratio $\epsilon_{\mu\mu}/\epsilon_{sb}\gtrsim 21\,(11)$. Again we put in $\epsilon_{\mu\mu}=$ sin$^2\,\theta_e\leq 1$, it then leads to the upper bound $\epsilon_{sb}\lesssim 0.05\,(0.09)$, which is also consistent with the corresponding value in the CKM matrix.

\subsubsection{Lepton Flavor Violation}

\begin{figure}[t]
\centering
\includegraphics[width=0.8\linewidth]{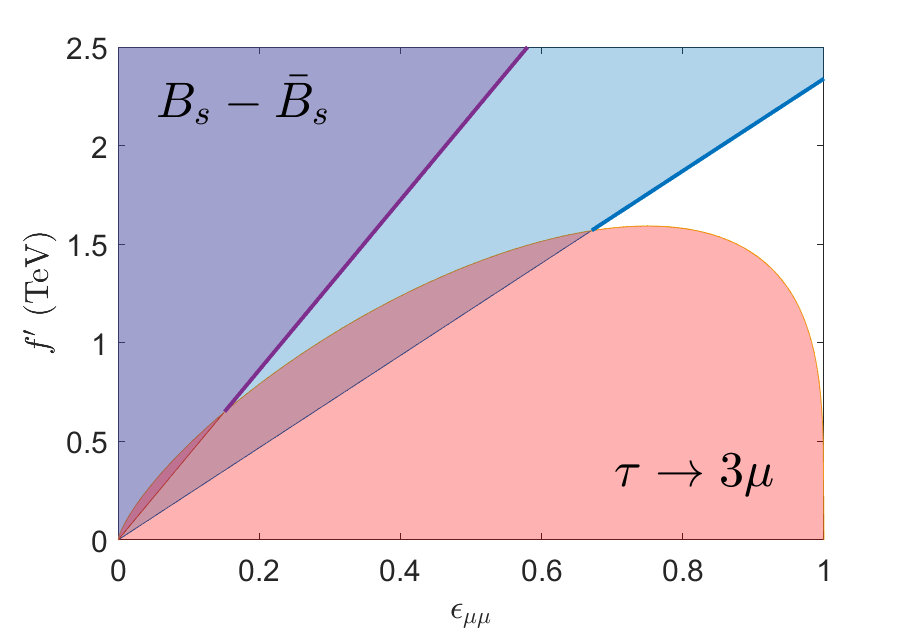}
\caption{The experimental constraints in the plane of $f'$ vs. $\epsilon_{\mu\mu}$. The shaded region is excluded by the corresponding measurements with $C_{LL}=-0.70$ (blue) and $C_{LL}=-0.38$ (purple). The strong blue and purple line labels the boundary of the available parameter space in the two different cases, which are important when we discuss the dimuon direct searches.}
\label{LFV}
\end{figure}

In the lepton sector, there are also strong constraints from flavor changing neutral currents. The off-diagonal term in the lepton charge matrix $Q_{e_L}$ will introduce lepton flavor violating decays. Because we only specified the rotation between the second and the third generation, the most important LFV effects show up in $\tau$ decays, especially $\tau\to 3\mu$. The relevant term generated by the $Z'$ boson in the effective Lagrangian is given by
\begin{equation}
\Delta\mathcal{L}_{\text{LFV}}=\frac{g_{Z'}^2}{16M_{Z'}^2}\text{sin}^3\,\theta_e \,\text{cos}\,\theta_e (\bar{\tau}_L\gamma^\rho\mu_L)(\bar{\mu}_L\gamma_\rho\mu_L)~,
\end{equation}
which leads to a branching ratio
\begin{align}
BR(\tau\to 3\mu)&=\frac{2m_\tau^5}{1536\pi^3\Gamma_\tau}\left(\frac{g_{Z'}^2}{16M_{Z'}^2}\text{sin}^3\,\theta_e \,\text{cos}\,\theta_e\right)^2\nonumber\\
&=1.28\times10^{-6}\,\left(\frac{1\text{ TeV}}{f'}\right)^4\epsilon_{\mu\mu}^3(1-\epsilon_{\mu\mu})~.
\end{align}
The value should be $<2.1\times 10^{-8}$ at $90\%$ CL \cite{Hayasaka:2010np}, which requires
\begin{align}
\left(\frac{1\text{ TeV}}{f'}\right)^4\epsilon_{\mu\mu}^3(1-\epsilon_{\mu\mu})< 1.6\times10^{-2}~.
\end{align}
It puts a strong constraint on the available parameter space, especially around the middle value of $\epsilon_{\mu\mu}$. The exclusion plot combining the requirement of neutral current B anomalies and the constraint from $B_s-\bar{B}_s$ mixing in the plane of $f'$ vs. $\epsilon_{\mu\mu}$ is shown in figure~\ref{LFV}.

The strong blue and purple lines correspond to the lower bound of $\epsilon_{\mu\mu}$ with the values $C_{LL}=-0.70$ and $C_{LL}=-0.38$ respectively, as shown in eq.~\eqref{minmumu}. Therefore, the white region is the allowed parameter space with $C_{LL}=-0.70$, where the small $\epsilon_{\mu\mu}$ region is totally excluded with a minimal value $\epsilon_{\mu\mu}\geq 0.67$. In this parameter space, the largest $Z'$ coupling in the lepton sector is $Z'\mu\mu$ and thus we only need to focus on the $\mu\mu$ channel when searching for $Z'$ in the LHC. On the other hand, the blue shaded region, which is a viable parameter space when $C_{LL}=-0.38$, allows a minimum $\epsilon_{\mu\mu}\geq 0.15$. In this region, the $\tau\tau$ and $\mu\tau$ final states will also play important roles in the direct searches. Therefore, to separate the two different conditions, we will call them ``Large angle region''(white) and ``Small angle region''(blue) in the following discussion.

Theoretically, the limit $\epsilon_{\mu\mu}=1$ implies that the muon should be the real third generation lepton in the flavor basis. Although we only have a little understanding about the origin of SM flavor structure, it is still unnatural to have a light muon be the third generation lepton instead of a heavy tau. From this point of view, the small $\epsilon_{\mu\mu}$ region is preferred and therefore worth more studies, which is not covered in the previous work \cite{Chung:2021ekz}.

%\newpage

\subsection{Direct $Z'$ Searches}\label{sec:Collider}

The measurements of flavor physics in the last section put constraints on the mixings and the scale $f'=M_{Z'}/g_{Z'}$. Direct searches, on the other hand, can provide the lower bound on the mass $M_{Z'}$ directly. A general $Z'$ collider search has been discussed in \cite{Allanach:2019mfl}, and in this section, we will focus on the parameter space provided by our model.

\subsubsection{Production cross section}

In the model, the $Z'$ boson only couples to the third generation fermions in the flavor basis. Even after rotating to the mass basis, the couplings to the first and second generation quarks are still suppressed due to the small mixing angles. Therefore, the dominant production comes from the process $b\bar{b}\to Z'$. In the following discussion, we will ignore all the other production processes and the small mixing angle $\theta_d$. In this way, the cross section is given by
\begin{equation}
{\sigma}(b\bar{b}\to Z') \equiv \frac{g_{Z'}^2}{16}\cdot \sigma_{bb}(M_{Z'}),
\end{equation}
where the coupling dependence is taken out. The $\sigma_{bb}$ (as a function of $M_{Z'}$) is determined by the bottom-quark parton distribution functions \cite{Martin:2009iq, Alwall:2014hca}.

\subsubsection{Decay width and branching ratios}

The partial width of the $Z'$ boson decaying into Weyl fermion pairs $\bar{f_i}f_j$ is given by
\begin{equation}
\Gamma_{ij}=\frac{C}{24\pi}g_{ij}^2M_{Z'},
\end{equation}
where $g_{ij}$ is the coupling of $Z'\bar{f_i}f_j$ vertex and $C$ counts the color degree of freedom. In the limit that all $m_f$ are negligible, we get the total relative width as
\begin{equation}
\frac{\Gamma_{Z'}}{M_{Z'}}=\frac{1}{24\pi}g_{Z'}^2\sim 1.3\,\%\cdot g_{Z'}^2~.
\end{equation}
We can see the narrow width approximation is only valid up to $g_{Z'}\sim 2.7$. If $g_{Z'}$ is larger than that, the direct search results need to be revised.

The dominant decay channels are the diquarks channel of the third generation with
\begin{equation}
Br(t\bar{t})\sim Br(b\bar{b}) \sim 37.5\%~.
\end{equation}
Decays to the light quarks and exotic decays like $tc$ and $bs$ are also allowed but strongly suppressed due to the small mixing angles in quark sectors. The main constraint is expected to come from the clear dilepton channels. Based on the specified mixing matrices we gave, the branching ratios are
\begin{align}
&Br(\tau\tau) \sim 6.25~(1+(1-\epsilon_{\mu\mu})^2)~\%~, \\
Br(\ell^+\ell^-) \sim 12.5\% \quad \implies \quad
&Br(\mu\tau) \sim 12.5~\epsilon_{\mu\mu}(1-\epsilon_{\mu\mu})~\%~, \\
&Br(\mu\mu) \sim 6.25~\epsilon_{\mu\mu}^2~\%~.
\end{align}
Notice that we only rotate the left-handed leptons, so there is always $\tau\tau$ decay from the right-handed side. In the large angle region, the bound $\epsilon_{\mu\mu}\geq 0.67$ from the flavor constraints implies $Br(\mu\mu)\geq 3 \%$. In this case, the $\mu\mu$ final state turns out to be the most promising channel and puts the stringent constraint on the $M_{Z'}$. On the other hand, in the small angle region, $\tau\tau$ and $\mu\tau$ could instead be the discovery channels.

\subsubsection{The $\mu\mu$ channel search}

\begin{figure}[t]
\centering
\includegraphics[width=0.7\linewidth]{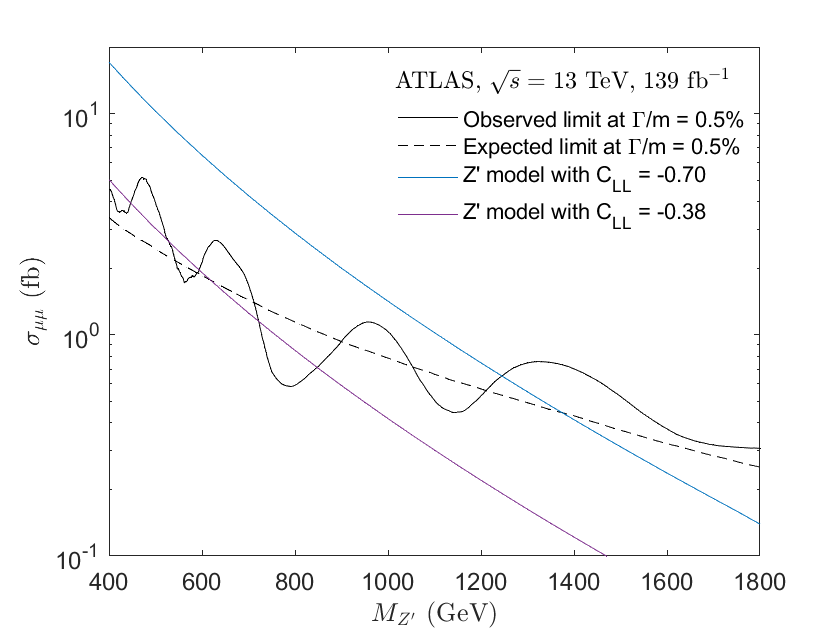}
\caption{Upper limits at 95\% CL on the cross section times branching ratio $\sigma_{\mu\mu}$ as a function of $M_{Z'}$ for 0.5\% relative width signals for the dimuon channel (black). Observed limits are shown as a solid line and expected limits as a dashed line. Also shown are theoretical predictions of the minimal cross section for $Z'$ in the model assuming $C_{LL}=-0.70$ (blue)  and $-0.38$ (purple).}
\label{mumu}
\end{figure}

From the production cross section and the branching ratios we got, we can calculate the cross section for the dimuon final state
\begin{equation}
\sigma_{\mu\mu} \equiv \sigma\times Br(\mu\mu) = \frac{1}{256}\,\sigma_{bb}\cdot g_{Z'}^2~\epsilon_{\mu\mu}^2.
\end{equation}
To analyze the constraints from the dimuon channel, we rewrite the cross section as
\begin{equation}
\sigma_{\mu\mu} = \frac{1}{256}\,\sigma_{bb}\cdot g_{Z'}^2\,\epsilon_{\mu\mu}^2
=\frac{1}{256}\,\sigma_{bb}\cdot g_{Z'}^2f'^2 \times\left(\frac{\epsilon_{\mu\mu}}{f'}\right)^2
=\frac{1}{256}\,\sigma_{bb} \cdot M_{Z'}^2 \times \left(\frac{\epsilon_{\mu\mu}}{f'}\right)^2,
\end{equation}
where the mass $M_{Z'}$ and the scale $f'$ are both in the unit of TeV. The rewrite allows us to separate the $M_{Z'}$ dependence part and the rest of the part only depends on $\epsilon_{\mu\mu}$ and $f'$. Looking back to the figure~\ref{LFV}, the ratio $\epsilon_{\mu\mu}/f'$ is the slope of a straight line in the plot. Then, from the $B_s-\bar{B}_s$ mixing measurement, we know the upper bound on the value of the ratio as shown in eq. \eqref{minmumu}, which gives
\begin{equation}
\sigma_{\mu\mu} \geq \frac{1}{256}\,\sigma_{bb}\cdot \left(\frac{M_{Z'}}{1 \text{ TeV}}\right)^2\times\left(\frac{1}{2.3(4.3) \text{ TeV}}\right)^2.
\end{equation}
The equality holds when the value sits at the strong blue (purple) line in figure \ref{LFV}. It gives the minimal cross section as a function of $M_{Z'}$ that allows us to compare with the experimental results. The current best search comes from the ATLAS \cite{ATLAS:2019erb} with an integrated luminosity of 139 fb$^{-1}$. The result is shown in figure~\ref{mumu}.

Notice that the bound by collider searches also weakly depends on the relative width ${\Gamma_{Z'}}/{M_{Z'}}$. In figure \ref{mumu} we choose a narrow relative width $\sim 0.5\%$, which corresponds to $g_{Z'}\sim 0.6$. The value is roughly consistent with the ratio between the bound on the $M_{Z'}$ and the scale on the corresponding segment in figure \ref{LFV}.

From the plot, the current bound for the large angle region is $M_{Z'}\gtrsim 1250$ GeV. For the small angle region, which has $\sim 3.5$ times smaller dimuon cross section, the $M_{Z'}$ could be as low as $400$ GeV. The most important message from the plot is: we are now exploring the most relevant parameter space in this $Z'$ model. Therefore, small changes in the input parameter can change the bound by a lot. Also, the increase of sensitivity in the near future will test the rest of the relevant parameter space.

\subsubsection{The $\tau\tau$ channel search}

\begin{figure}[t]
\centering
\includegraphics[width=1.0\linewidth]{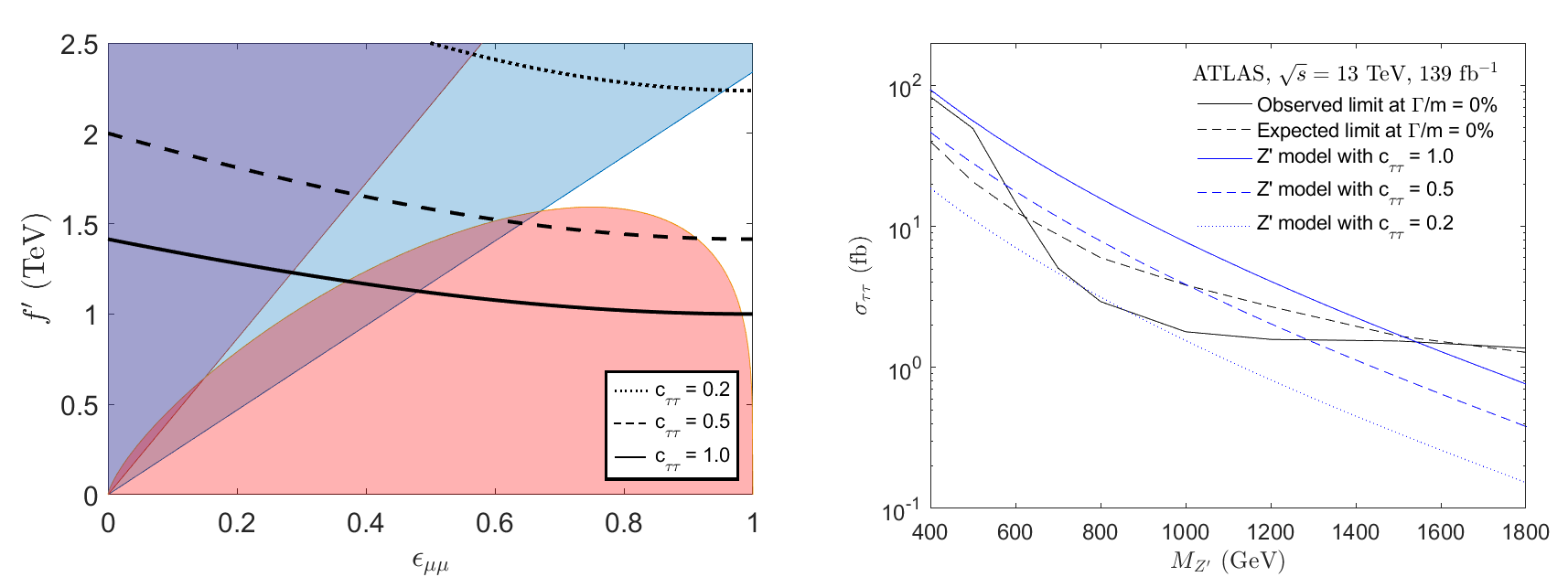}
\caption{Left: The contour with constant $c_{\tau\tau}(\text{TeV}^{-2})=1.0\,$(solid) $,0.5\,$(dashed), and $0.2\,$(dotted). Right: Upper limits at 95\% CL on the cross section times branching ratio $\sigma_{\tau\tau}$ as a function of $M_{Z'}$ for 0.5\% relative width signals for the ditau channel (black). Observed limits are shown as a solid line and expected limits as a dashed line. Also shown are theoretical predictions of the minimal cross section for $Z'$ in the model assuming different $c_{\tau\tau}(\text{TeV}^{-2})$.}
\label{tautau}
\end{figure}

For the large angle region, the only relevant direct search is the $\mu\mu$ final state. However, once we look into the parameter space with smaller $\epsilon_{\mu\mu}$ allowed by a small $C_{LL}$, the $\tau\tau$ channel will also matter. From the production cross section and the branching ratios we got, we can derive the cross section for the $\tau\tau$ final state
\begin{equation}
\sigma_{\tau\tau} \equiv \sigma\times Br(\tau\tau) = \frac{1}{256}\,\sigma_{bb}\cdot g_{Z'}^2~(1+(1-\epsilon_{\mu\mu})^2)~.
\end{equation}
To analyze the constraint from the $\tau\tau$ channel, we again take out the $M_{Z'}$ dependence by rewriting the cross section as
\begin{equation}
\sigma_{\tau\tau} = \frac{1}{256}\,\sigma_{bb}\cdot g_{Z'}^2f'^2 \times
\left(\frac{1+(1-\epsilon_{\mu\mu})^2}{f'^2}\right)=\frac{1}{256}\,\sigma_{bb}\cdot M_{Z'}^2 \times c_{\tau\tau}~,
\end{equation}
where the factor
\begin{equation}
c_{\tau\tau}\equiv \frac{1+(1-\epsilon_{\mu\mu})^2}{f'^2} ~.
\end{equation}
Then we can put curves with $c_{\tau\tau}$ equal to some constants and calculate the corresponding $\sigma_{\tau\tau}$ as a function of $M_{Z'}$. It allows us to compare with the current experimental results, which we take from the ATLAS \cite{ATLAS:2020zms} with an integrated luminosity of 139 fb$^{-1}$ as shown in figure~\ref{tautau}. The allowed blue parameter space below the solid black line was strongly constrained with bound $M_{Z'}\gtrsim 1600$ GeV. The higher contour on the left implies a smaller $\sigma_{\tau\tau}$, above which the parameter space is still viable with a smaller $M_{Z'}$.

\subsubsection{The $\mu\tau$ channel search}

\begin{figure}[t]
\centering
\includegraphics[width=1.0\linewidth]{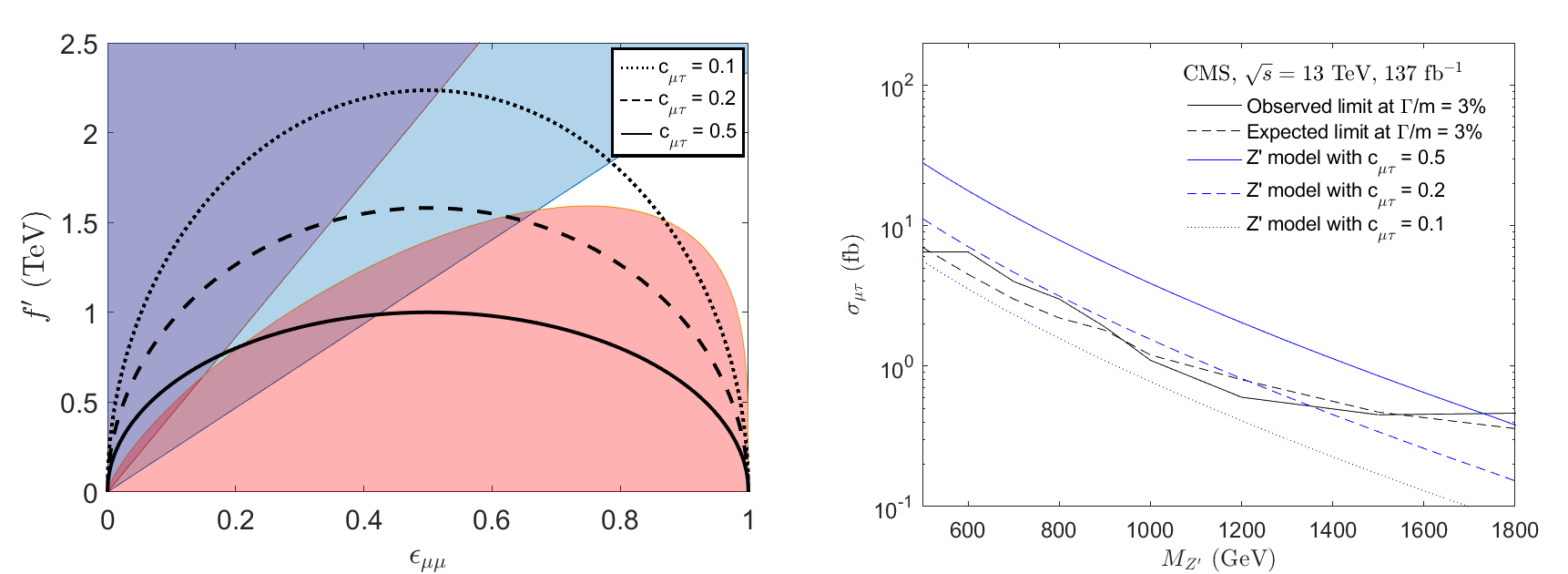}
\caption{Left: The contour with constant $c_{\mu\tau}(\text{TeV}^{-2})=0.5\,$(solid) $,0.2\,$(dashed), and $0.1\,$(dotted). Right: Upper limits at 95\% CL on the cross section times branching ratio $\sigma_{\mu\tau}$ as a function of $M_{Z'}$ for 3\% relative width signals for the $\mu\tau$ channel (black). Observed limits are shown as a solid line and expected limits as a dashed line. Also shown are theoretical predictions of the minimal cross section for $Z'$ in the model assuming different $c_{\mu\tau}(\text{TeV}^{-2})$.}
\label{mutau}
\end{figure}

Besides the $\tau\tau$ channel, in the small angle region, the lepton flavor violating channel is also important. The cross section for $\mu\tau$ final state is given by
\begin{equation}
\sigma_{\mu\tau} \equiv \sigma\times Br(\mu\tau) = \frac{1}{256}\,\sigma_{bb}\cdot g_{Z'}^2~(2\epsilon_{\mu\mu}(1-\epsilon_{\mu\mu}))~.
\end{equation}
Again, we can rewrite the cross section as
\begin{equation}
\sigma_{\mu\tau}= \frac{1}{256}\,\sigma_{bb}\cdot g_{Z'}^2f'^2 \times
\left(\frac{2\,\epsilon_{\mu\mu}(1-\epsilon_{\mu\mu})}{f'^2}\right)= \frac{1}{256}\,\sigma_{bb}\cdot M_{Z'}^2 \times ~c_{\mu\tau}~.
\end{equation}
where the factor
\begin{equation}
c_{\mu\tau}\equiv\frac{2\,\epsilon_{\mu\mu}(1-\epsilon_{\mu\mu})}{f'^2}~.
\end{equation}
The contour of constant $c_{\mu\tau}$ is given on the left of figure~\ref{mutau}. The corresponding $\sigma_{\mu\tau}$ as a function of $M_{Z'}$ is derived and compared with the current experimental results, which we get from the CMS \cite{CMS:2021tau} with an integrated luminosity of 137 fb$^{-1}$ as shown on the right of figure~\ref{mutau}. The left end of the allowed blue parameter space is again strongly constrained with bound $M_{Z'}\gtrsim 1800$ GeV. On the other hand, the region above the dotted contour is weakly constrained from the direct search in the ${\mu\tau}$ channel. A comprehensive discussion considering all the channels together will be presented in the next subsection.

%\newpage
\subsubsection{Other decay channels}

Other decay channels are unlikely to be the discovery channel. However, if the $Z'$ boson is discovered, the next thing to do will be to look for the same resonance in other channels. Through the searches, we can decide the partial widths and figure out the couplings of the $Z'$ boson to other SM fields. The structure of couplings can help us distinguish between different $Z'$ models, which can truly reveal the nature of the B anomalies. In our model, the $Z'$ boson couples universally to all the third generation SM fermions in the flavor basis. After the transformation to the mass basis, the universality no longer exists. However, a unique partial width ratio will be preserved as
\begin{equation}
\Gamma_{tt}:\Gamma_{bb}:\Gamma_{\ell\ell}:\Gamma_{\nu\nu}\sim 3:3:1:1,
\end{equation}
where $\Gamma_{\ell\ell}$ is the sum of all the charged lepton partial widths. This ratio is an important prediction of our $Z'$ model. Moreover, since an anomaly-free $U(1)'$ within SM fermions can not achieve this ratio, it can be indirect evidence for the existence of additional fermions beyond the SM, such as hyperfermions in our model.

Other final states with light quarks and electrons are in principle allowed. However, they are expected to be strongly suppressed due to the small rotation angles. Only when a huge amount of $Z'$ are produced can we manage to see the signal in these channels. Once they could be measured, they will provide new observables for us to probe the rotation matrices between the flavor basis and mass basis of SM fermions. Unlike the charge current mediated by $W$ boson in the SM, which can only tell us the difference between $U_{u_L}$ and $U_{d_L}$, i.e. CKM matrix, the neutral current mediated by a third-generation-philic $Z'$ boson can provide more information about the individual matrix. What can be measured is the product of matrix $U_{F_{L/R}}^{33}$ defined as (under standard parameterization)
\begin{equation}
U_{F_{L/R}}^{33}\equiv
U_{F_{L/R}}^\dagger    
\begin{pmatrix}
0   &  0  &  0    \\
0   &  0  &  0    \\
0   &  0  &  1    \\
\end{pmatrix}
U_{F_{L/R}}=
\begin{pmatrix}
s_{13}^2  &  s_{13}c_{13}s_{23}e^{-i\delta}  &  s_{13}c_{13}c_{23}e^{-i\delta}    \\
s_{13}c_{13}s_{23}e^{-i\delta}   &  c_{13}^2s_{23}^2  &  c_{13}^2s_{23}c_{23}    \\
s_{13}c_{13}c_{23}e^{-i\delta}   &  c_{13}^2s_{23}c_{23}     &  c_{13}^2c_{23}^2    \\
\end{pmatrix} ,
\end{equation}
which is a symmetric matrix. The $33$ in the superscript implies that it is from the rotation of a unit vector in the $(3,3)$ direction. Therefore, it is not sensitive to $\theta_{12}$. But still, if we can measure the other angles, $\theta_{13}$ and $\theta_{23}$, and the phase $\delta$, that will be great progress toward understanding the original Yukawa matrices and thus the origin of Yukawa couplings.

\subsection{Combined Analysis}\label{sec:Combine}

In this study, we are interested in the bounds on the scale $f'$, which is related to the symmetry breaking scale $f$, and the mass $M_{Z'}$, which is important for collider searches. Using the relation we derived in the last section, we can calculate the $\sigma_{\mu\mu}$, $\sigma_{\mu\tau}$, and $\sigma_{\tau\tau}$ for the corresponding parameter space in figure~\ref{LFV}. Combined with the direct search results from ATLAS and CMS, we can get the allowed parameter space in the $f'$ vs. $M_{Z'}$ plane. However, for the different regions of parameter space, the analysis of phenomenology is quite different so we separate them into the large angle region and small angle region.

\subsubsection{Large angle region}

\begin{figure}[t]
\centering
\includegraphics[width=0.7\linewidth]{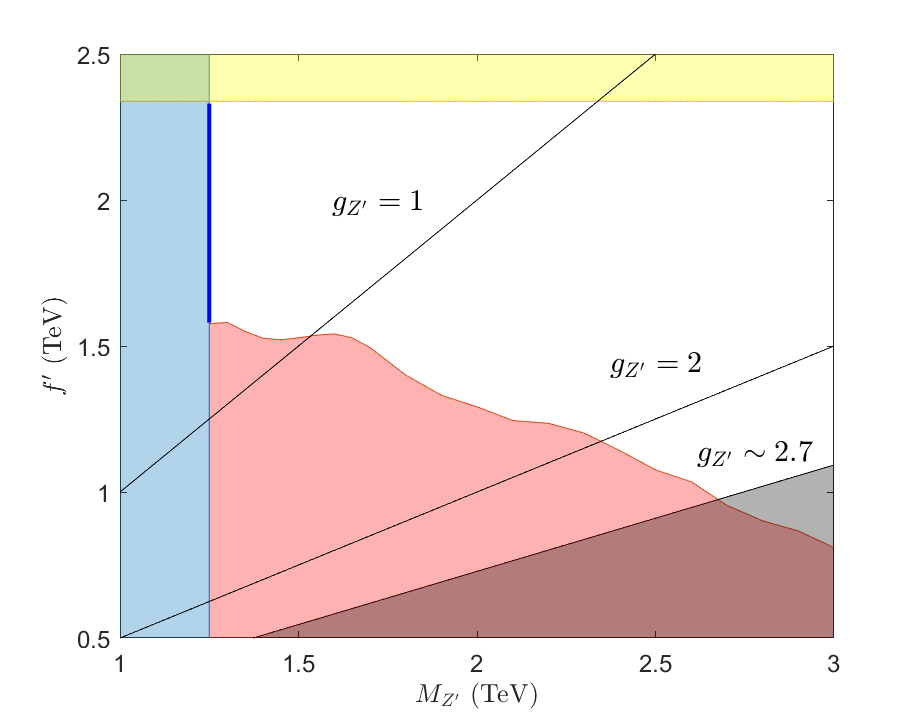}
\caption{Constraints in the $f'$ vs. $M_{Z'}$ plane for $M_{Z'}$ below $3$ TeV (for the large angle region). The white region is currently allowed, where $\epsilon_{\mu\mu}$ and $\epsilon_{sb}$ can be chosen to satisfy \eqref{Banomaly} from the requirement of the B anomalies. The shaded regions are excluded by the corresponding constraints in figure~\ref{LFV} combined with the direct searches, where we use the ATLAS 139 fb$^{-1}$ dimuon searches. The yellow region is the upper bound in the large angle region as derived in eq.~\eqref{minmumu}. The three black straight lines represent different values of $g_{Z'}$. In the black shaded region, the large $g_{Z'}$ violates the narrow width approximation and the direct searches result is not valid.}
\label{fvsMzL}
\end{figure}

Start with the large angle region. In this parameter space, we only need to consider the constraint from the dimuon searches. The resulting parameter space is shown in figure~\ref{fvsMzL}. The blue region is excluded by the $B_s-\bar{B}_s$ meson mixing, which gives the lower bound $M_{Z'}\gtrsim 1.25$ TeV as we have shown in figure~\ref{mumu}. The strong blue line corresponds to the same parameter space as in figure~\ref{LFV} with $M_{Z'}\sim 1.25$ TeV. The yellow region, also excluded by the $B_s-\bar{B}_s$ meson mixing, sets the maximum value for $f'$ as shown in eq.~\eqref{minmumu}. Once the stronger constraint from $B_s-\bar{B}_s$ meson mixing is placed, the yellow line will move downward, and the blue line will move rightward. The red region, which is excluded by $\tau\to 3\mu$, restricts the space from below. It places the lower bound on $f'$, which will be pushed upward if the constraint becomes stronger. We can also see the data fluctuations in the dimuon search become the fluctuations on the red curve.

The strength of the coupling $g_{Z'}$ of three different values is labeled as the black straight lines in the plot, where $g_{Z'}\sim 2.7$ also sets the bound above which the narrow width approximation is not valid. The direct searches in the near future will extend both blue and red exclusion regions rightward, which will cover all the points around the boundary. The lower bound on $M_{Z'}$ will be pushed to 2 TeV, and most of the interesting parameter space will be explored during the HL-LHC era \cite{ATL-PHYS-PUB-2018-044, CidVidal:2018eel}.

Notice that the black shaded region is not excluded by experiments, but requires other analysis to set up the bound for such a wide resonance. Also, since we are talking about an $U(1)$ gauge symmetry, the strong coupling will quickly reach the Landau pole as discussed in section~\ref{sec:Strength}.

%\newpage

\subsubsection{Small angle region}

\begin{figure}[t]
\centering
\includegraphics[width=0.9\linewidth]{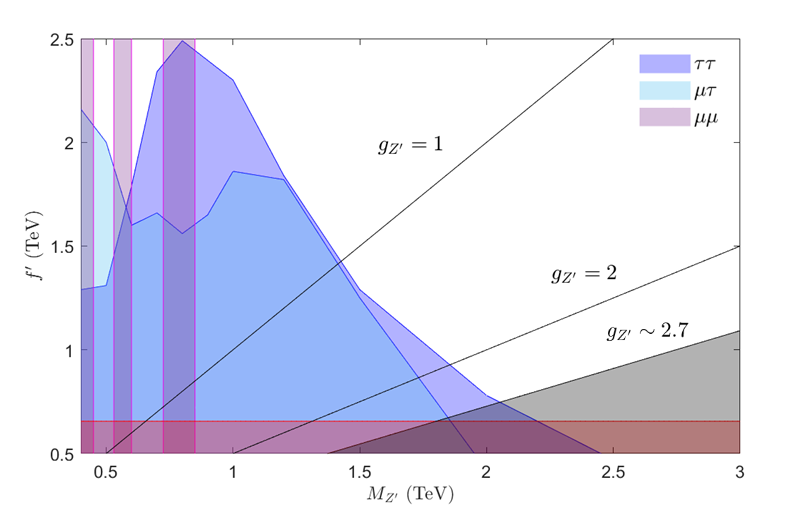}
\caption{Constraints in the $f'$ vs. $M_{Z'}$ plane for $M_{Z'}$ below $3$ TeV (for the small angle region). The white region is currently allowed, where $\epsilon_{\mu\mu}$ and $\epsilon_{sb}$ can be chosen to satisfy \eqref{Banomaly} from the requirement of the B anomalies. The shaded regions are excluded by the corresponding constraints from the different direct searches, including $\tau\tau$, $\mu\tau$, and $\mu\mu$ final states, combined with $B_s-\bar{B}_s$ constraint and NCBA requirement. The red shaded region is the lower bound due to the constraint from $\tau\to 3\mu$, which can be seen from the lower left corner of the blue region in figure \ref{LFV}. The three black straight lines represent different values of $g_{Z'}$. In the black shaded region, the large $g_{Z'}$ violates the narrow width approximation and the direct searches result is not valid.}
\label{fvsMzS}
\end{figure}

For the small angle region, the $\mu\tau$ and $\tau\tau$ channels should also be considered. To simplify the analysis, in this work we alone the boundary of the blue parameter space in figure~\ref{LFV}, where constraint from the $\mu\mu$ direct search is weakest as shown in figure~\ref{mumu}. The resulting parameter space is given in figure~\ref{fvsMzS}.

In this case, the bound from above is much weaker ($f' \leq 4.3$ TeV), which is not shown in the figure. Since we stick to the strong purple line, the constraint from $\tau\to 3\mu$ is only relevant at the bottom of the parameter space, which gives a lower bound on $f' \gtrsim 600$ GeV as shown in the figure \ref{LFV}. A larger parameter space is opened up for a TeV-scale $Z'$. Moreover, the sub-TeV $Z'$ is also possible in the gap which is not ruled out by the dimuon direct search. The shaded region will expend either from the stronger constraint on $B_s-\bar{B}_s$ meson mixing or the new direct search results from the LHC Run3.

\section{Connecting the B anomalies with the Hierarchy problem}\label{sec:Combine}

The analysis in section~\ref{sec:Z'} is purely based on the $Z'$ interactions with the SM fermions, which allow us to get the constraints on its scale $f'$ and mass $M_{Z'}$ from flavor physics and direct searches. In other words, it is just a story of a $Z'$ with interaction terms given in eq. \eqref{Zint1}. However, we want to connect the $Z'$ boson to its origin - the $SU(4)/Sp(4)$ FCHM. The connection will import constraints from Higgs and electroweak physics as well as some theoretically preferred bounds on the coupling and the scale. The new constraints will be added to figure~\ref{fvsMzL} and figure~\ref{fvsMzS} and be discussed one by one in this section.

To connect the constraints from different sectors, we need to relate their scales first. In section~\ref{sec:U(1)}, we have shown that, for a $Z'$ in the $SU(4)/Sp(4)$ FCHM, its scale $f'$ is related to the symmetry breaking scale (the real scale in the UV model) $f$ by $f'\approx (2/N_{HC})f$, which depends on the number of hypercolor $N_{HC}$. In the following discussion, we display two specific cases  
\begin{equation}
(1)~N_{HC}=2:\quad f \approx f',\qquad (2)~N_{HC}=4:\quad f \approx 2f'~,
\end{equation}
where the first case corresponds to $SU(2)$ strongly coupled gauge interaction and the second case corresponds to $Sp(4)$.

For the new constraints, we start with the coupling $g_{Z'}$. In FCHMs, the $Z'$ boson also couples to the hyperfermions as shown in eq. \eqref{Zint0}. Knowing all the fermions that participate in the $U(1)'$ interaction, we can calculate the RG running as we have done in section~\ref{sec:Strength}. The value of coupling $g_{Z'}$, which will reach the Landau pole around the Planck scale, is marked by the black dotted line. The value that will reach the Landau pole before $10^3$ TeV is considered so strong that the $g_{Z'}$ above it is disfavored theoretically. Since there is already an upper bound on $g_{Z'}$ from the requirement of narrow width approximation but with a larger value compared to the new bound from the issue of the Landau pole, we use the new bound labeled by the black dashed line and mark the exclusion region by the black region. No matter which bound we use, it is not an experimental constraint. The solution with large $g_{Z'}$ is possible but requires beyond narrow width approximation and clarification for the UV completion, which are challenging.

Next, there are also constraints on the scale $f$. We have discussed two experimental constraints in section~\ref{sec:HiggsEW}. For the electroweak precision tests, it is hard to include all the effects and determine the exact bound, so no constraints will be applied from the measurements. On the other hand, the Higgs coupling directly measures the nonlinearity of the Higgs boson. Therefore, we can exclude the scale $f$ below the lower bound $780$ GeV derived in eq. \eqref{Higgsbound}, which is shown as a purple shaded region in the following figures. Besides the experimental constraints, we do not want the fine-tuning in the Higgs potential to be too large. Although it does not really rule out any parameter space, it marks the region that is considered unnatural and theoretically disfavored. The bound of fine-tuning is hard to decide. In section~\ref{sec:Tuning}, we have taken
$8\pi\sqrt{-\alpha} \sim 1.6$ TeV as the fine-tuning upper bound. Therefore, $f\geq1.6$ TeV is labeled by the green shaded region.

After combining all the bounds, the experimentally and theoretically favored parameter space are again shown with the two separate regions.

\subsection{Large angle region}

\begin{figure}[t]
\centering
\includegraphics[width=1.0\linewidth]{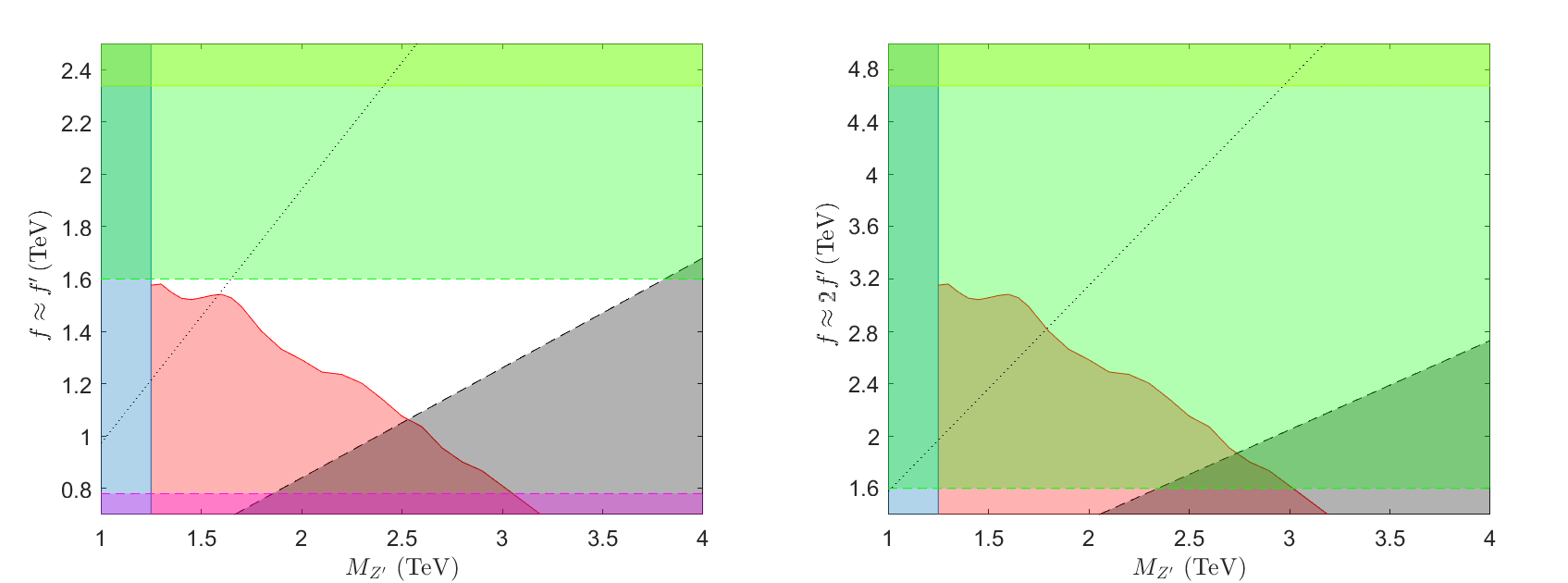}
\caption{Constraints in the $f(f')$ vs. $M_{Z'}$ plane for $N_{HC}=2$ (Left) and $N_{HC}=4$ (Right). The yellow, blue, and red regions are the same as in figure~\ref{fvsMzL}. The dotted black line represents the coupling that reaches a Landau pole at the Planck scale. The shaded black region below the dashed black line labels the coupling which reaches a Landau pole below $10^3$ TeV. The horizontal magenta shaded region is excluded by the Higgs coupling measurements. The horizontal green shaded region is theoretically disfavored due to the fine-tuning issue in the Higgs potential.}
\label{fvsMzNL}
\end{figure}

For the large angle case, we consider the two different $N_{HC}$. The allowed parameter space is shown in figure~\ref{fvsMzNL}. We found that the black region, which is the upper bound of the theoretically favored $g_{Z'}$, combined with the red region, which is the combination of the constraints from $\tau\to 3\mu$ and direct $Z'$ searches, already gives a lower bound on the scale $f\gtrsim 1$ TeV for $N_{HC}=2$ ($f\gtrsim 2$ TeV for $N_{HC}=4$), which is already stronger than the bound from Higgs measurement labeled by the purple shaded region. The requirement of naturalness (green) also applies a stronger bound compared to the experimental constraints from $B_s-\bar{B}_s$ neutral meson mixing (yellow).

In figure~\ref{fvsMzNL} (left), the $N_{HC}=2$ scenario, even though more than half of the original allowed parameter space is now considered fine-tunned, there is still some viable parameter space, which will be explored in the near future. On the other hand, in figure~\ref{fvsMzNL} (right), we find that the whole experimentally viable parameter space in the $N_{HC}=4$ scenario is disfavored theoretically. As already shown in figure~\ref{fvsMzL}, the experimentally favored parameter space (excluding the strong coupling region) requires the scale $f' \gtrsim 1$ TeV. Therefore, the lower bound on the scale $f$ will easily surpass the natural parameter space when $N_{HC}>3$. Since the $SU(4)/Sp(4)$ symmetry breaking pattern can only be realized by $SP(N_{HC})$ hypercolor group with even number $N_{HC}$, the only choice is $N_{HC}=2$.

Back to figure~\ref{fvsMzNL} (left), most of the viable parameter space is located between the black dotted line and the black dashed line. That means, if it is the case, it will point to a new scale between the TeV scale and the Planck scale, which is closely related to the origin of this $U(1)'$ symmetry.

\subsection{Small angle region}

\begin{figure}[t]
\centering
\includegraphics[width=0.8\linewidth]{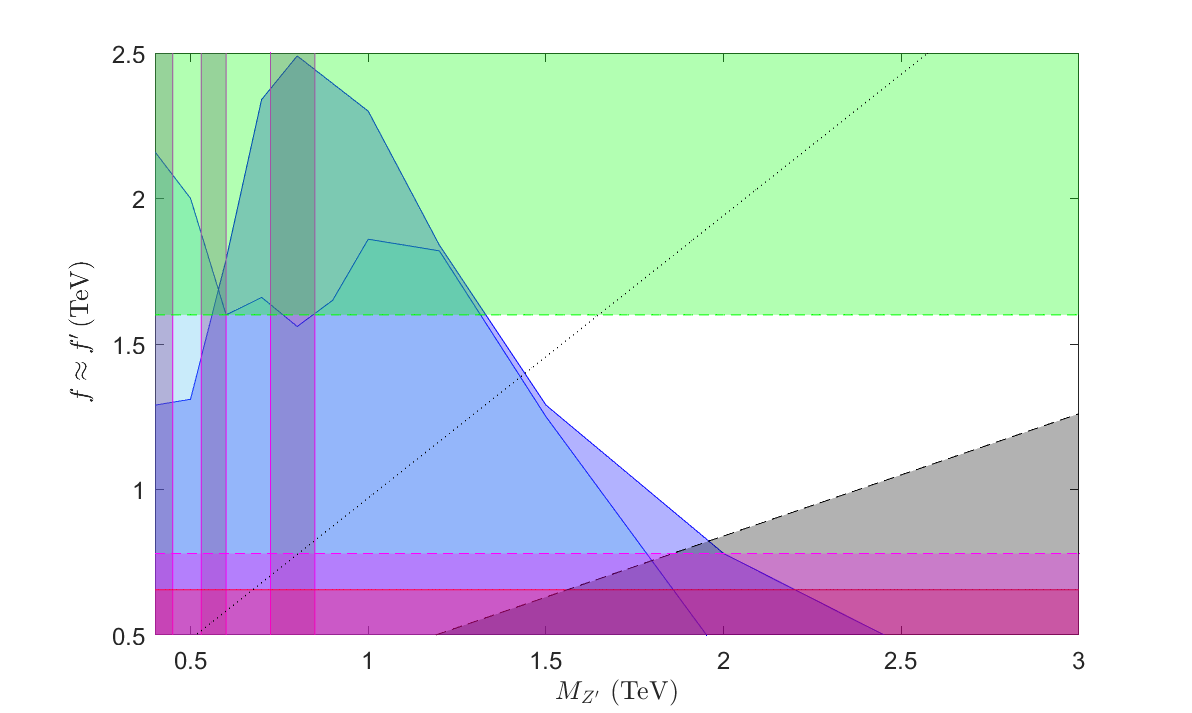}
\caption{Constraints in the $f(f')$ vs. $M_{Z'}$ plane for $N_{HC}=2$ (for the small angle region). The shaded regions in figure~\ref{fvsMzS} are also shown here. The dotted black line represents the coupling that reaches a Landau pole at the Planck scale. The shaded black region below the dashed black line labels the coupling which reaches a Landau pole below $10^3$ TeV. The horizontal magenta shaded region is excluded by the Higgs coupling measurements. The horizontal green shaded region is theoretically disfavored due to the fine-tuning issue in the Higgs potential.}
\label{fvsMzNS}
\end{figure}

In the small angle region, we only consider the $N_{HC}=2$ case because the $N_{HC}=4$ scenario is also disfavored as in the large angle region. The allowed parameter space is shown in figure~\ref{fvsMzNS}. It shares very similar properties as the large angle case. Most of the allowed parameter space is disfavored now, but still, there is viable parameter space with $M_{Z'}>1.2$ TeV. Moreover, in the sub-TeV region, the constraints from different channels are very close to each other and might still be viable. If that is the case, the $Z'$ boson will be within the reach of LHC Run3.

\section{Conclusions}\label{sec:Conclusion}

The naturalness principle is eager for new particles around the TeV scale. However, none of them have been observed in the LHC so far. We claim that the current anomalies in B-meson semileptonic decays might be the first hint of the new particle. From another point of view, the neutral current B anomalies also seek a solution. Its scale is so close to the TeV scale that we can not help but wonder if it shares a common origin with electroweak symmetry breaking and the Higgs boson. In either view, we should look for a model aiming at addressing the two problems together. With this motivation, we presented the minimal model connecting the B anomalies with the hierarchy problem, which is based on a $SU(4)/SP(4)$ fundamental composite Higgs model with an additional TeV-scale $Z'$ boson from the $U(1)_{SM_3-HB}$ gauge symmetry.

A composite Higgs model is an appealing solution to the hierarchy problem, and the fundamental CHM with a fermionic UV completion is even more attractive. However, additional pNGBs other than the Higgs doublet are always introduced and should be the leading contribution to the BSM effects. 
One of the possible BSM effects observed by the LHC is the neutral current B anomalies, which can be explained by a $Z'$ boson. The spectrum with order `` Higgs - $Z'$ - Others '' is wanted, and we find that such spectrum can be nicely generated in a $SU(4)/Sp(4)$ FCHM if the symmetry is gauged.

It turns out that the symmetry gauged should be $U(1)_{SM_3-HB}$ in order to explain the neutral current B anomalies. The $Z'$ boson, which only couples to the third generation SM fermions in the flavor basis, can naturally introduce FCNCs with CKM suppression. The parameter space that can explain the NCBAs is shown in figure~\ref{LFV}. Also, it is naturally beyond the current direct searches because of its small cross section, which is mainly through the $b\bar{b}$ fusion. The current bound $M_{Z'}\gtrsim 1$ TeV is still around the expectation and leaves interesting allowed parameter space shown in figure~\ref{fvsMzL} and figure~\ref{fvsMzS}. Even lighter $Z'$ is possible and will soon be tested in the LHC Run3.

In fact, from previous studies in $Z'$ solutions, it is not a surprise that the $Z'$ in our model can also get rid of all the experimental constraints. What is novel in this study is that viable parameter space still exists after considering the constraints from $SU(4)/Sp(4)$ FCHM, which include Higgs physics and several theoretical preferences. In the end, the $SU(4)/SP(4)$ FCHM with $N_{HC}=2$ is favored as shown in section \ref{sec:Combine}. The failed scenario with $N_{HC}=4$ also serves as a good example showing that success is nontrivial.

The property that the $Z'$ boson only couples to the third generation is the key for this idea to work. Such a TeV-scale third-generation-philic $Z'$ boson can naturally escape from direct searches in the first two runs of the LHC. However, the origin of this property is unknown (some ideas are presented in the appendix) and it might relate to the mass hierarchy between the third generation and the other two. The complete model should try to include the explanation, which is left to a future study.

\acknowledgments

I thank Hsin-Chia Cheng, Markus Luty, and Cheng-Wei Chiang for many useful discussions. I am also grateful to Ben Allanach and Wolfgang Altmannshofer for reading the previous paper and giving many helpful suggestions. This work is supported by the Department of Energy Grant number DE-SC-0009999.

%\newpage

\appendix

\section{The origin of the $U(1)'$ gauge symmetry}\label{sec:Origin}

In section \ref{sec:U(1)'}, we get the charge assignment of $U(1)'$ from the requirement of anomaly cancellation, which ends up with the $U(1)_{SM_3-HB}$ gauge symmetry. However, it is put in by hand without further explanation. In this appendix, we try to bring up some possible explanations for the origin of this charge assignment, including the form of $SM-HB$ and the generation dependence.

\subsection{The $U(1)'$ subgroup from the extended hypercolor group}\label{sec:EHC}

Analogous to the $U(1)_{B-L}$ symmetry, which can arise as the $U(1)$ subgroup of the Pati-Salam unified group \cite{Pati:1974yy}, the $U(1)_{SM-HB}$ can also arise from a large unified group. The group should contain both the SM gauge group and hypercolor group. Also, it should allow the SM fermions and hyperfermions to stay in the same representation. This type of group is called the extended hypercolor group (EHC). Historically, the idea has been used in the Extended Technicolor models \cite{Dimopoulos:1979es, Eichten:1979ah} to introduce interaction between the SM fermions and technifermions, which can generate the SM fermion masses from the technifermion condensate. Recently, the ideas have been applied in the composite Higgs models \cite{Cacciapaglia:2019dsq, Cacciapaglia:2020jvj}, to provide a UV-complete CHM valid up to the Planck scale.

With the extended hypercolor group, where $G_{EHC}\supset G_{SM}\times G_{HC}$, after $G_{EHC}$ is broken at some scale $\Lambda_{EHC}$, the remaining subgroups could include, besides $G_{SM}\times G_{HC}$, an $U(1)_{EHC}$ that satisfies the form of $SM-HB$ as we expected. To display how it works, we show a concrete example combining the third generation SM fermions with hyperfermions. To combine them in the same representation, we can first unify the SM gauge group $G_{SM_3}$ to the Pati-Salam group $G_{PS_3}=SU(4)_{PS_3}\times SU(2)_L\times SU(2)_R$, where the subscript 3 means only the third generation fermions are included. The third generation SM fermions and hyperfermions transform under $SU(4)_{PS_3}\times SP(N_{HC}) \times SU(2)_L\times SU(2)_R$ as
\begin{align}
&F_L=(4,1,2,1),\quad \psi_L=(1,N_{HC},2,1),\nonumber\\
&F_R=(\bar{4},1,1,2),\quad \psi_R=(1,N_{HC},1,2).
\end{align}
A similar structure in the $SU(2)_L\times SU(2)_R$ sector can be apparently found. The simplest extension for $G_{EHC_3}$ would be $SU(4+N_{HC})_{EHC_3} \times SU(2)_L\times SU(2)_R$, where fermions are unified in the same representations for each chirality as
\begin{equation}
f_{L/R}=
\begin{pmatrix}
t^r   &  t^g  &  t^b  &  \nu_\tau   & U^1 & \cdots  &  U^{N_{HC}}  \\
b^r   &  b^g  &  b^b  &  \tau      & D^1  & \cdots  &  D^{N_{HC}}  \\
\end{pmatrix}_{L/R}~.
\label{fermion}
\end{equation}
The desired $U(1)_{EHC_3}$ subgroup is the one with a generator given by
\begin{equation}
Y'=c_{Y'}\text{ Diag}(1,1,1,1,-{4}/{N_{HC}},\cdots -{4}/{N_{HC}})~.
\end{equation}
The charge under $U(1)_{EHC_3}$ are the same as the $U(1)'=U(1)_{SM_3-HB}$ we discussed in this paper if $c_{Y'}=1/4$. The breaking of $G_{EHC_3}$ is designed such that $U(1)_{EHC_3}$ is preserved until the TeV scale and then broken by the hyperfermions condensate. The $U(1)_{EHC_3}$ symmetry originating from an anomaly-free non-abelian gauge group will also be free from the quantum anomaly.

%\newpage

\subsection{The generation-dependence from the horizontal group}\label{sec:Gen}

So far, we did not explain why the $Z'$ only couples to the third generation. One possible way is to combine the flavor-blind $U(1)_{SM-HB}$ with another flavor-dependent gauge group, such as a horizontal group. In this way, the remaining $U(1)'$ symmetry could be the linear combination of $U(1)_{SM-HB}$ and $U(1)_{H}$, which is a subgroup of horizontal symmetry. The method has been applied on $U(1)_{B-L}$ \cite{Alonso:2017bff, Alonso:2017uky} and successfully gives an anomaly-free $U(1)'$ gauge symmetry with desired couplings to fermions. 

The desired flavor-dependent $U(1)_H$ should preserve $U(2)$ symmetry among the first and second generations. Also, it needs to be anomaly-free. The only choice is given by the $U(1)_H$ with a generator $c_H\text{ Diag}(-1,-1,2)$ operating on the $F=(F_1,F_2,F_3)$. $c_H$ is an arbitrary normalization factor.\footnote{One can also view $U(1)_H$ as a subgroup of a larger non-abelian horizontal symmetry, such as $SU(3)_H$. Then, $c_H$ should be normalized as $1/\sqrt{12}$} Under the $U(1)_H$ symmetry, the third (first and second) generation SM fermions carry charge $2c_H(-c_H)$. Then, the desired $U(1)'$ can be the linear combination of $U(1)_{SM-HB}$ and $U(1)_H$ with $c_H=1/12$ with charges shown in table \ref{tab:charge1}. It successfully reproduces the same charge assignment as the $U(1)_{SM_3-HB}$ gauge symmetry we studied in this paper.
 
\begin{table}[t]
\centering
\begin{tabular}{|c|c|c|c|}
\hline
& SM$_{1,2}$ & SM$_{3}$ & HF\\ \hline
$Q_{SM-HB}$ & $1/12$ & $1/12$ & $-1/N_{HC}$ \\ \hline
$Q_{H}$ & $-1/12$ & $1/6$ & $0$ \\ \hline
$Q'=Q_{SM-HB}+Q_{H}$ & $0$ & $1/4$ & $-1/N_{HC}$ \\ \hline
\end{tabular}
\caption{The charges of SM fermions and HF hyperfermions under different $U(1)$ symmetry. The subscript of the fermions denotes the generation. The $Q'$ is the sum of $Q_{SM-HB}$ and $Q_{H}$.\label{tab:charge1}}
\end{table}

\begin{table}[t]
\centering
\begin{tabular}{|c|c|c|c|c|}
\hline
& SM$_{1,2}$ & SM$_{3}$ & HF$_{1,2}$ & HF$_3$ \\ \hline
$Q_{EHC}$ & $1/12$ & $1/12$ & $-1/(3 \,N_{HC})$ & $-1/(3 \,N_{HC})$ \\ \hline
$Q_{H}$ & $-1/12$ & $1/6$ & $-1/12$ & $1/6$ \\ \hline
$Q'=Q_{EHC}+Q_{H}$ & 0 & $1/4$ & $-1/12-1/(3 \,N_{HC})$ & $1/6-1/(3 \,N_{HC})$ \\  \hline
\end{tabular}
\caption{The charges of SM fermions and HF hyperfermions under different $U(1)$ symmetry. The subscript of the fermions denotes the generation. The $Q'$ is the sum of $Q_{EHC}$ and $Q_{H}$.\label{tab:charge2}}
\end{table}

Combining with the idea from the extended hypercolor in Appendix \ref{sec:EHC}, another attractive scenario arises, where hyperfermions are also charged under horizontal symmetry. In this case, both SM fermions and hyperfermions are complete multiplet of flavor-blind $G_{EHC}$ and flavor-dependent $G_{H}$. The $U(1)'$ is the linear combination of $U(1)_{EHC}$ with $c_{Y'}=1/12$ and $U(1)_H$ with $c_H=1/12$ as shown in table \ref{tab:charge2}. The new $U(1)'$ can also naturally decouple from the first and second generation SM fermions. Moreover, the $U(1)'$ can be viewed as the remnant of the broken $G_{EHC}\times G_H$, which is like the hypercharge $U(1)_Y$ as the remnant of the broken $SU(4)_{PS}\times SU(2)_R$ in the Pati-Salam model. The $Z'$ model has the same phenomenology due to the same interaction with SM fermions. However, the hyperfermion sector is extended. Three sets of hyperfermions will generate more Higgs doublets, which make the Higgs sector more complicated. Also, their condensates will all contribute to the $Z'$ mass, but the resulting mass $M_{Z'}$ is similar to what we get in eq.~\eqref{Zmass} because the sum of the hyperfermion charges ($=-1/N_{HC}$) remains unchanged. Additional hyperfermions will also affect the RG running of the $U(1)'$ coupling. The detail of this setup requires another complete study.

%\newpage
\bibliographystyle{jhepbst}
\bibliography{Flavon_Ref}{}

\end{document}